\documentclass[twocolumn,preprint]{aastex62}
\usepackage[utf8x]{inputenc}
\usepackage{amsmath,amsfonts,amssymb,mathrsfs,bm,enumerate,graphicx}
\usepackage[T1]{fontenc}
\usepackage[all]{hypcap}
\hypersetup{
    colorlinks=true,
    linkcolor=magenta,
    filecolor=magenta,      
    urlcolor=blue,
}

\usepackage{ulem}
\usepackage{threeparttable}
\usepackage{ae,aecompl}
\usepackage{wasysym}

\def\red#1 {\textcolor{red}{#1}\ }   %using this command will put the text in red, so as to be easily seen.
\def\blue#1 {\textcolor{blue}{#1}\ }   %using this command will put the text in blue, so as to be easily seen.
\def\purple#1 {\textcolor{purple}{#1}\ }   %using this command will put the text in blue, so as to be easily seen.

\newcommand{\D}{{\rm d}}

\shorttitle{Evection Resonance in AGN Disks}
\shortauthors{Mu\~noz et al.}

\begin{document}

\title{\bf Eccentric Mergers of Intermediate-Mass Black Holes from   Evection Resonances in AGN Disks}
\author[0000-0003-2186-234X]{Diego J.\ Mu\~noz}
\affiliation{Center for Interdisciplinary Exploration and Research in Astrophysics (CIERA)
and
Department of Physics and Astronomy
Northwestern University, 
1800 Sherman Avenue, Evanston, IL 60208, USA}
\affiliation{Facultad de Ingenier\'ia y Ciencias, Universidad Adolfo Ib\'a\~nez, Av.\ Diagonal las Torres 2640, Pe\~nalol\'en, Santiago, Chile}
\affiliation{Millennium Institute for Astrophysics, Chile}
\author[0000-0002-4337-9458]{Nicholas S.\ Stone}
\affiliation{Racah Institute of Physics, The Hebrew University, Jerusalem, 91904, Israel}
\author[0000-0003-0412-9314]{Cristobal Petrovich}
\affiliation{Pontificia Universidad Cat\'olica de Chile, Facultad de F\'isica, Instituto de Astrof\'sica, Av.\ Vicu\~na Mackenna 4860, 7820436 Macul, Santiago, Chile 5 Millennium Institute for Astrophysics, Chile}
\affiliation{Millennium Institute for Astrophysics, Chile}
\author[0000-0002-7132-418X]{Frederic A.\ Rasio}
\affiliation{Center for Interdisciplinary Exploration and Research in Astrophysics (CIERA)
and Department of Physics and Astronomy Northwestern University, 
1800 Sherman Avenue, Evanston, IL 60208, USA}

\begin{abstract}
We apply the theory of nonlinear resonance capture to the problem of
a black hole binary (BHB) orbiting a supermassive black hole (SMBH) while embedded in the accretion disk of an active galactic nucleus (AGN).  If successful, resonance capture can trigger dramatic growth in the BHB eccentricity, with important consequences for the BHB merger timescale as well as for the gravitational wave (GW) signature such an eccentric merger may produce.
This resonance capture may occur
when the orbital period around the SMBH (the ``outer binary'')  and
the apsidal precession of the BHB  (the ``inner  binary'')  are in a 1:1 commensurability.
This effect is analogous to the phenomenon of lunar evection resonance in the early Sun-Earth-Moon system, with the distinction that in the present case, the BHB apsidal precession is due to general relativity, rather than rotationally-induced distortion. In contrast to the case of lunar evection, however, the BHB (inner binary) also undergoes
orbital decay driven by GW emission, rather than expansion driven by tidal dissipation. This distinction fundamentally alters the three-body dynamics, forbidding resonance capture, and limiting eccentricity growth. However, if the BHB migrates through of a gaseous AGN disk, 
the change in the outer binary can counterbalance the suppressing effect of BHB decay, permitting
evection resonance capture and the production of eccentric BHB mergers. We compute the likelihood of resonance capture assuming an agnostic distribution of parameters for the three bodies involved and for the properties of the AGN disk. We find that intermediate-mass ratio BHBs (involving an intermediate-mass black hole and a stellar-mass black hole) are the most likely to be captured into evection resonance and thus undergo an eccentric merger. We also compute the GW signature of these mergers, showing that they can enter the LISA band while eccentric.
\end{abstract}

\keywords{Astrophysical black holes (98);
Supermassive black holes (1663);
Active galactic nuclei (16);
Gravitational wave soumrces (677);
Astrodynamics (76);
Orbital resonances (1181)}

%%%%%%%%%%%%%%%%%%%%%%%%%
\section{Introduction}
The recent detection of the LIGO-Virgo event GW190521  \citep[LIGO Scientific Collaboration \& Virgo Collaboration,][]{LIGO2020} appears to have finally confirmed 
 the existence of intermediate-mass black holes (IMBHs; those with masses in the range $M_\bullet\sim10^2-10^5M_\odot$). After 
decades in which the existence of IMBHs was supported only by indirect observational evidence \citep[see, e.g.,][]{vanderm2004,mez17,greene20}, by contested dynamical modeling \citep{noyo08, vanderm10}, or by model-dependent accretion disk analysis \citep{farr09}, gravitational waves (GWs) 
have now provided a definitive smoking gun for compact objects in this mass range. If confirmed by future observations, a significant IMBH population would revolutionize the field of compact object astrophysics, with profound ramifications for dense star cluster dynamics \citep{gual09}, SMBH formation \citep{volo10}, and even topics in fundamental physics like cosmological large-scale structure \citep{mada01} or the existence of ultra-light bosons \citep{wen21}.

Mergers involving an IMBH have long been recognized as promising sources of GWs for ground-based and space-based observatories \citep{mill02a,mill02b,mill03,mill04,mill09,amar07,amar18a,amar18b}.  These include both mergers  between an IMBH and a stellar-mass black hole, and those between an IMBH and a supermassive black hole.  Both of these intriguing possibilities are termed ``intermediate mass ratio inspirals'' (or IMRIs), in contrast to both the comparable-mass mergers detected so far by the LIGO-Virgo-KAGRA (LVK) collaboration and the extreme mass ratio inspirals expected to be found by future space-based detectors such as {\it LISA}.  
These two categories of IMRIs could be detected with either LVK or {\it LISA}, respectively \citep{amar07,mand08,amar18b},
or by third generation facilities like the Einstein Telescope.  Any IMRI detection would be doubly valuable: firstly, in providing hard-to-come-by information on the demographics of IMBHs. Secondly, these IMRIs are expected
to probe gravity in the strong-field regime \citep{amar07, rodr12}.  As tests of general relativity, they have both unique advantages \citep{yune10} and unique disadvantages \citep{mand09} in comparison to more standard GW signals.

The most frequently explored routes to forming IMBHs are direct collapse of gas in high-redshift, low-metallicity environments \citep{loeb94, brom03}, growth via collisions in dense stellar clusters \citep[e.g.,][]{bahc75,mill02a, omuk08}, and the top-heavy mass function thought to describe Pop III stars \citep{hira14}.
A novel alternative channel, however, is that of
massive object formation in AGN disks  \citep{good04,mcker12}.

AGN disks have also been proposed as efficient hotbeds for compact binary formation and GW events  \citep{ston17}. Indeed, recent years have seen many studies exploring the variety of GW events that may emerge from a population of compact object binaries embedded in an AGN disk \citep[e.g.,][]{mcker18,taga20,sams20}. Most of these studies, however, focus on stellar mass black hole binaries (BHBs), without devoting much attention to more massive objects. This is in spite of some theoretical arguments that AGN disks could harbor IMBHs formed by either gravitational collapse \citep{good04} or via hierarchical mergers \citep{yang19}.

Moreover, AGN disks have been identified as possible sites for {\it eccentric} BHB mergers \citep[e.g.,][]{sams20,taga21}. But once again, most of these studies focus primarily on stellar mass binaries and their GW signatures in the LIGO band. In general, most studies focused on eccentric GW sources entering the LVK band are, in one shape or another, based on the dynamical assembly of BHBs via GW capture, which can ``initialize'' binaries in-band with
non-negligible eccentricities \citep{bena02,kocs06,olea09,kocs12,gond18,sams18}.
While alternative eccentricity-pumping mechanisms -- such as the von Zeipel-Lidov-Kozai (ZLK) effect -- have been studied for BHBs orbiting a SMBH \citep{liu18}, it is not clear that the high inclinations needed for ZLK to operate can be found in BHB-SMBH triples embedded in gaseous AGN disk.
In this work, we present an alternative eccentricity-pumping mechanism that differs from GW capture and  the ZLK mechanism. This mechanism, known from lunar theory as the {\it evection resonance}, is able to increase the BHB eccentricity to high values, and does not require an initially inclined BHB orbit in order to operate.  We shall see, however, that the evection resonance works best when applied to the merger of a stellar-mass BH and an IMBH: an IMRI.
\subsection{The Phenomenon of Evection}
In modern celestial mechanics, evection refers to the term in the lunar disturbing function
proportional to $\cos(2\lambda_\odot-2\varpi_{\leftmoon})$, where $\lambda_\odot$ is the Sun's mean longitude and  $\varpi_{\leftmoon}$ is the Moon's longitude of pericenter \citep[e.g.,][$\S$ XII]{brouwer1961}. Being a short-period oscillatory term, this contribution is often
``averaged out''  in the secular approximation. When retained in the disturbing function,
this evection term usually only modulates
 $\varpi_{\leftmoon}$ and the eccentricity $e_{\leftmoon}$ in an oscillatory fashion. However, this forcing can translate into net eccentricity growth if $\cos2(\lambda_\odot-\varpi_{\leftmoon})$ is slowly evolving. Indeed, when
$\dot{\lambda}_\odot\approx\dot{\varpi}_{\leftmoon}$, the system is said to be in a state of 
``evection resonance'' \citep{tou98}, a phenomenon that is thought to have played an important role in the early evolution of the Earth-Moon system.

%%%%%%%%%%%%%%%%%%%%%%%%%%%%%%%%
%%%%%%%%%%%%%%%%%%%%%%%%%%%%%%%%
\begin{figure*}
\centering
\includegraphics[width=0.95\textwidth]{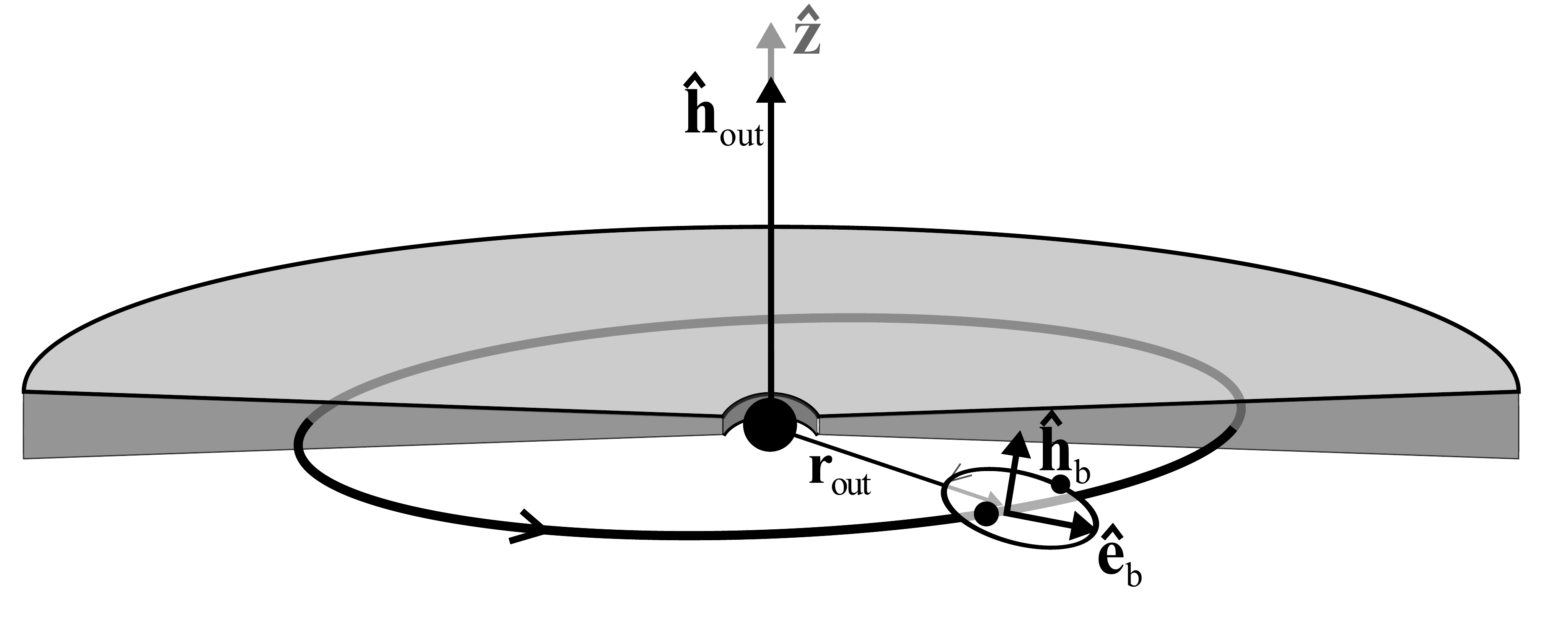}
\caption{Schematic representation of a three-body system
comprised of a compact BHB embedded in an AGN disk around a SMBH.
The compact binary's orbital phase is averaged out from the Hamiltonian ${\cal H}^*$ (Equation~\ref{eq:averaged_hamiltonian}), but the wide orbit's true anomaly $\nu_{\rm out}$ appears explicitly in ${\cal H}^*$ 
and in the equations of motion. Consequently,
the compact (``inner'') binary is entirely described by the Milankovitch
state vectors ${\bf h}_{\rm b}=L_{\rm b}\sqrt{1-e_{\rm b}^2}\hat{\bf h}_{\rm b}$ and ${\bf e}_{\rm b}=e_{\rm b}\hat{\bf u}_{\rm b}$ \citep{mil39}, which evolve in time according to Equations~\ref{eq:eom_milankovitch}. The orbital elements of the wide orbit
are assumed constant, except for the semi-major axis
$a_{\rm out}$, which decreases owing to, e.g., Type I disk migration.
\label{fig:orbits}}
\end{figure*}
%%%%%%%%%%%%%%%%%%%%%%%%%%%%%%%%
%%%%%%%%%%%%%%%%%%%%%%%%%%%%%%%%

Phenomena
akin to evection may occur in any hierarchical triple system in which no time average is carried out over the outer orbit \citep[e.g.,][]{tou15,spa16,xu16}.  In this work, we
extend the applicability of this effect to the dynamics of BHBs around a central SMBH.  In such case, the relevant commensurability is
\begin{equation}\label{eq:commensurability}
n_{\rm out}\approx \dot{\varpi}_{\rm b}
\end{equation}
where $n_{\rm out}$ is the mean motion (or orbital frequency) of the BHB around the SMBH and 
$ \dot{\varpi}_{\rm b}$ is the apsidal precession rate of the BHB, which is due to lowest-order
post-Newtonian corrections.

As with lunar evection, a commensurability like Equation~(\ref{eq:commensurability}) can be crossed
\citep[or ``swept''; e.g.,][]{war76} when the semi-major axes of the system slowly change in time.
But in contrast to lunar evection, the separation of the BHB (the ``smaller binary'') {\it decreases} due to gravitational wave radiation, instead of {\it growing} due to tidal dissipation. Another distinction from lunar evection is that the orbit of the BHB around the SMBH can decay due to nebular tides, which may also result in an evection commensurability  crossing \citep[e.g.,][]{spa16}.  Consequently,  evection commensurability crossing of BHBs  in 
AGN disks is governed by the combined effects of hardening and migration.

If commensurability crossing results in resonance capture, the three-body dynamics dictate that eccentricity can grow arbitrarily (other damping mechanism being absent), and thus the evolution of the BHB toward eventual merger can differ dramatically from its non-resonant counterpart.

In this work, we study BHBs in evection resonance.
In Section~\ref{sec:dynamics}, we overview the dynamics of BHBs around a SMBH as a hierarchical triple system, demonstrating that the singly-averaged system can be reduced to the classical second fundamental model of resonance.
In Section~\ref{sec:agn_disks}
describe how, and estimate how often, BHBs can be captured into an evection resonance,
estimating the gravitational wave signature of those systems that become resonant.
In Section~\ref{sec:discussion} we discuss the applications and limitations of our calculations.
Finally, In Section~\ref{sec:summary}, we summarize our findings.

%%%%%%%%%%%%%%%%%%%%%%%%%
\section{Black Hole Binaries orbiting a Super-Massive Black Hole}\label{sec:dynamics}
\subsection{Equations of Motion}\label{sec:equations}
The truncated Hamiltonian of a triple consisting
of a compact binary of total mass $M_{\rm b}{=}m_1{+}m_2$ and mass ratio $q_{\rm b}=m_2/m_1$ orbiting a SMBH of
mass $M_\bullet\gg M_{\rm b}$ is
\begin{equation}\label{eq:basic_hamiltonian}
\begin{split}
{\cal H} =& -\frac{{\cal G} m_1m_2}{2a_{\rm b}} -\frac{{\cal G}M_\bullet M_{\rm b}}{2a_{\rm out}}
+{\cal H}_{\rm 1PN}({\bf r}_{\rm b},\dot{{\bf r}}_{\rm b})\\
&-\frac{{\cal G}M_\bullet m_1m_2}{M_{\rm b}r_{\rm out}}
\frac{1}{2}\left[\frac{3({\bf r}_{\rm b}\cdot{\bf r}_{\rm out})^2}{r_{\rm out}^4}-\frac{r_{\rm b}^2}{r_{\rm out}^2}
\right]
\end{split}
\end{equation}
where ${\cal H}_{\rm 1PN}({\bf r}_{\rm b},\dot{\bf r}_{\rm b})$ is the first
post-Netownian correction to the two-body problem in Hamiltonian form \citep[e.g., see][]{straumann1984,sha18}.

We assume that both orbits are, to first order, Keplerian, and hence 
${\bf r}_{\rm b}$ and ${\bf r}_{\rm out}$ can be expressed in vectorial form as
\citep[e.g.,][]{tre14b}: ${\bf r}_i=r_i(\cos \nu_i \hat{\bf u}_i+\sin \nu_i \hat{\bf v}_i)$ for $i{=}{\rm b}{,}{\rm out}$.  The
$\nu_i$ are the true anomalies and the unit vectors $\hat{\bf u}_i=(u_x,u_y,u_z)$ and $\hat{\bf v}_i=(v_x,v_y,v_z)$
 define the orientation of each binary in space, with
$u_x=\cos\omega_i\cos\Omega_i-\cos I_i\sin\omega_i\sin\Omega_i$,
$u_y=\cos\omega_i\sin\Omega_i+\cos I_i\sin\omega_i\cos\Omega_i$,
$u_z=\sin I_i  \sin\omega_i$,
and
$v_x=-\sin\omega_i\cos\Omega_i-\cos I_i\cos\omega_i\sin\Omega_i$,
$v_y= -\sin\omega_i\sin\Omega_i+\cos I_i\cos\omega_i\cos\Omega_i$,
$v_z= \sin I_i  \cos\omega_i$.

Averaging over the period of the inner orbit \citep[e.g.,][]{nao13b,tre14b,liu15}, and discarding constant terms, we have
\begin{equation}\label{eq:averaged_hamiltonian}
\begin{split}
{\cal H}^* =&-\frac{3{\cal G}^2\mu_{\rm b}M_{\rm b}^2}{c^2a_{\rm b}^2\sqrt{1-e^2_{\rm b}}}
~~~~~~~~~~~~~~~~~~~~~~~~~~~~~~~~~~~~~~~~~~\\
&-{{\cal G}\mu_{\rm b} M_\bullet}\frac{a_{\rm b}^2}{4r_{\rm out}^3}
\big[1-6e_{\rm b}^2  -3L_{\rm b}^{-2}({\bf h}_{\rm b}\cdot\hat{\bf r}_{\rm out})^2 \\
&~~~~~~~~~~~~~~~~+15({\bf e}_{\rm b}\cdot\hat{\bf r}_{\rm out})^2
\big],\!\!\!
\end{split}
\end{equation}
where  $\mu_{\rm b}=m_1m_2/M_{\rm b}=q_{\rm b}(1+q_{\rm b})^{-2}M_{\rm b}$is the BHB's reduced mass.
The ``Milankovitch'' state vectors are the eccentricity or  Laplace-Runge-Lenz vector
${\bf e}_{\rm b}=e_{\rm b }\hat{\bf u}_{\rm b}$ and the angular momentum vector
${\bf h}_{\rm b}=L_{\rm b}\sqrt{1-e^2_{\rm b }}\hat{\bf h}_{\rm b}$,
where $L_{\rm b}=\mu_{\rm b}\sqrt{{\cal G}M_{\rm b}a_{\rm b}}$ and
$\hat{\bf h}_{\rm b}=\hat{\bf u}_{\rm b}\times \hat{\bf v}_{\rm b}$ (see orbit depiction in Figure~\ref{fig:orbits}).

  %%%%%%%%%%%%%%%%%%%%%%%%%%%%%%%%
%%%%%%%%%%%%%%%%%%%%%%%%%%%%%%%%
\begin{figure*}
\includegraphics[width=0.325\textwidth]{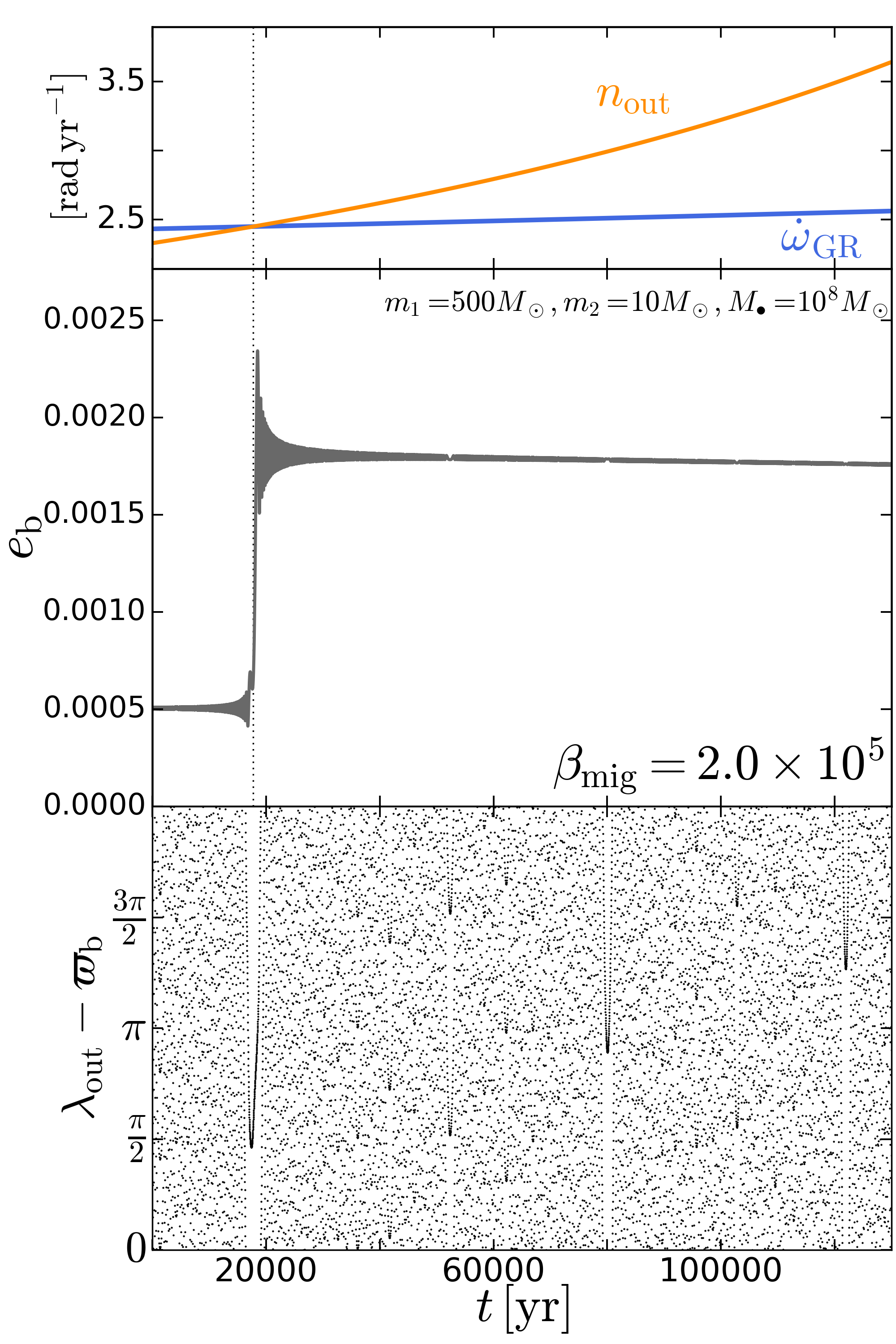}
\includegraphics[width=0.325\textwidth]{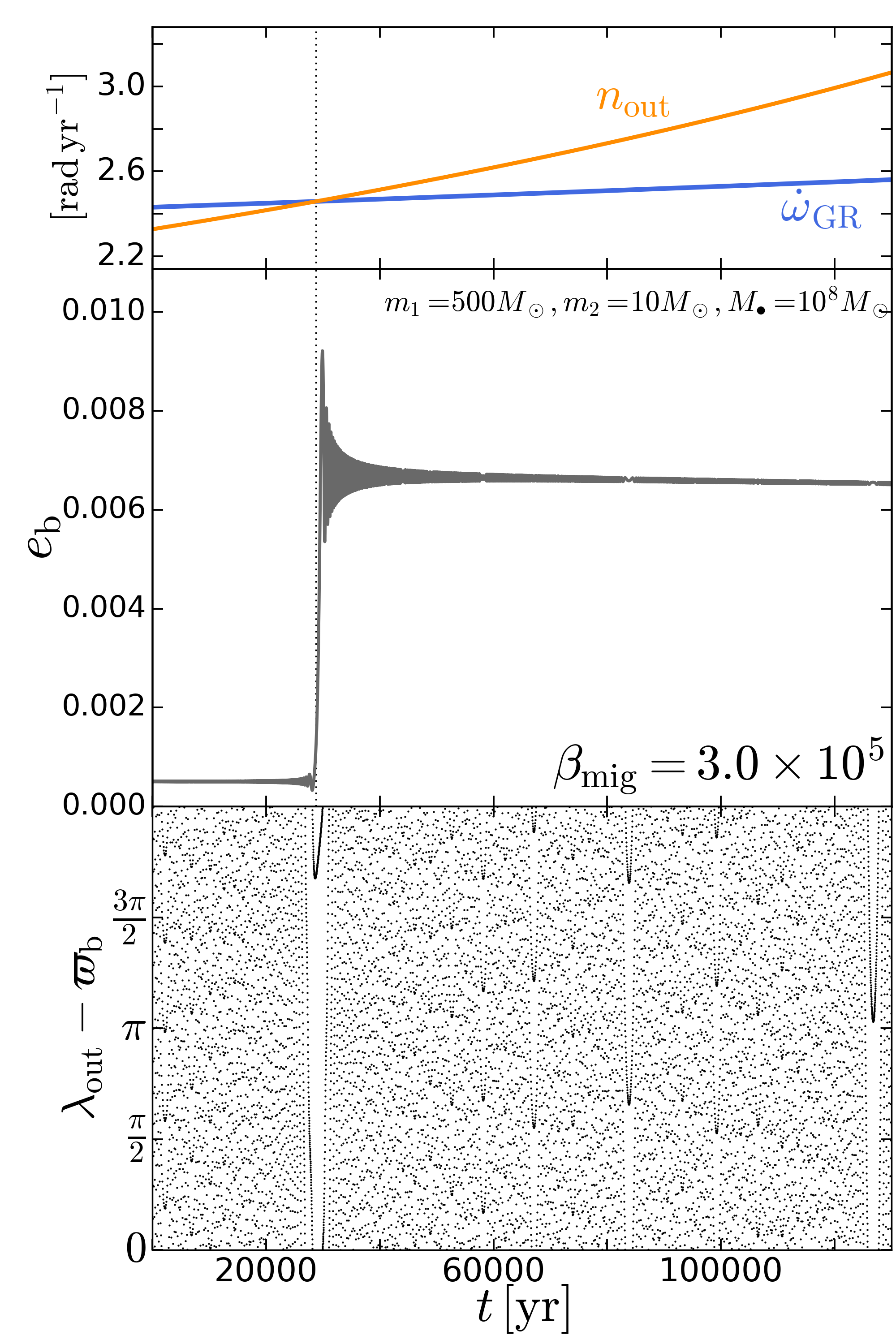}
\includegraphics[width=0.325\textwidth]{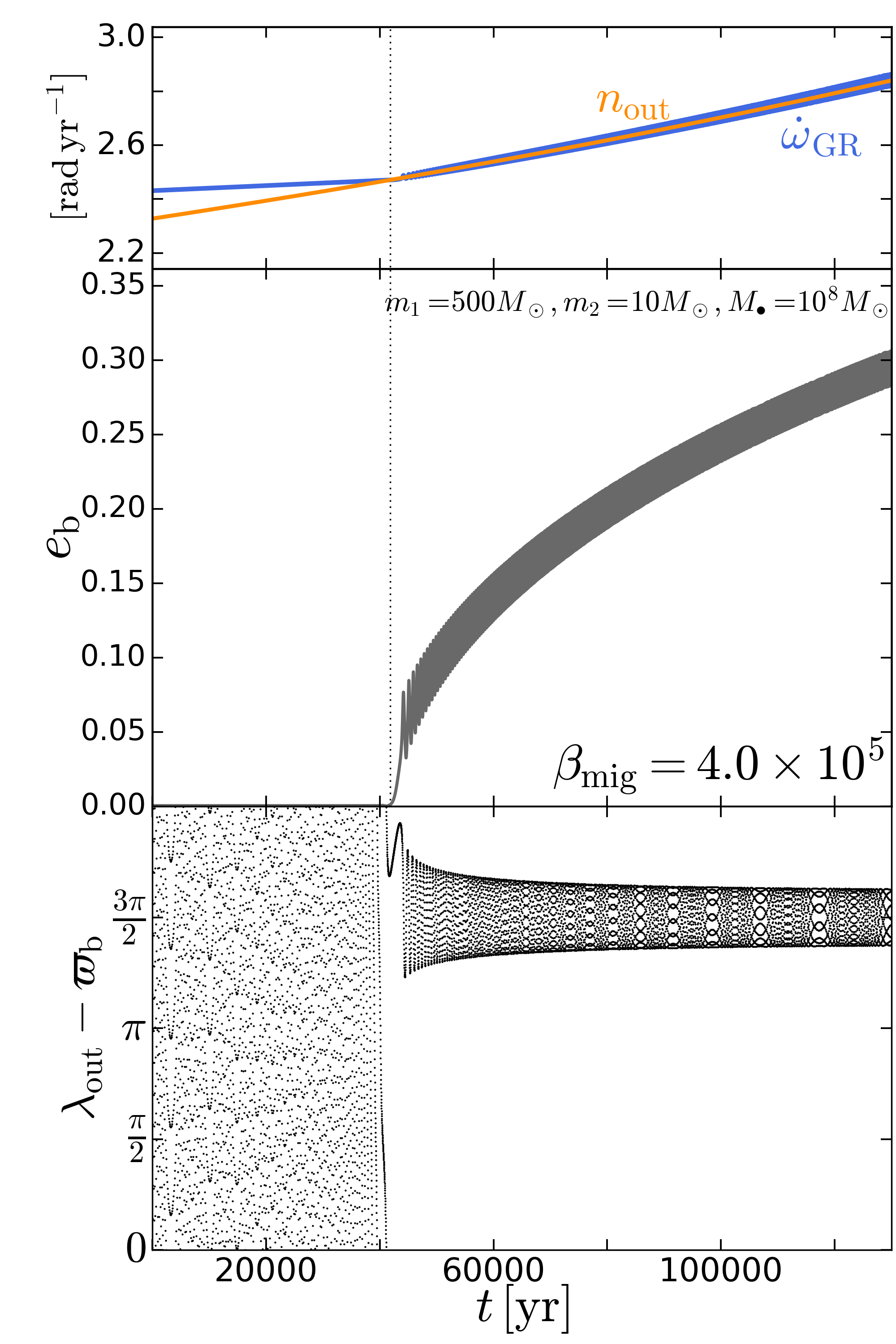}
\caption{Numerical solution of Equations~(\ref{eq:milankovitch_eom}) for a system with
$M_\bullet=10^8M_\odot$, $M_{\rm b}=510M_\odot$ $q_{\rm b}=0.02$,
$a_{\rm out,0}=900$~au and $a_{\rm b,0}=0.06$~au and for different values of the migration rate: $\beta_{\rm mig}=2\times10^5$,
$3\times10^5$ and $4\times10^5$ (left to right). In all panels, the commensurability
(Equation~\ref{eq:commensurability}) is crossed, with crossing times given by $t_\star=1.771\times10^4$~yr (left)
$t_\star=2.876\times10^4$~yr (middle)
 $t_\star=4.183\times10^4$~yr (right). While commensurability crossing is always accompanied by in a change in eccentricity only for 
  $\beta_{\rm mig}=4\times10^4$ that the change in $e_{\rm b}$ is substantial. The top panels illustrate how the frequencies
  $n_{\rm out}$ and $\dot{\omega}_{\rm GR}$ can remain commensurate
  for sufficiently high $\beta_{\rm mig}$. 
  Similarly, the bottom
  panels show that the canonical angle $\sigma$ can transition from circulating to librating in tandem with significant eccentricity growth.
\label{fig:eccentricity}}
\end{figure*}
%%%%%%%%%%%%%%%%%%%%%%%%%%%%%%%%
%%%%%%%%%%%%%%%%%%%%%%%%%%%%%%%%
% 

The  equations of motion derived from the single-average Hamiltonian~(\ref{eq:averaged_hamiltonian}) follow from the Poisson structure of  ${\cal H}^*$ \citep[e.g.,][]{tre14b}, and
are 
\begin{subequations}
\label{eq:eom_milankovitch}
\begin{align}
\begin{split}
\frac{\D{\bf e}_{\rm b}}{\D t}\bigg|_{\rm c}&=3n_{\rm b}
\frac{r_g}{a_{\rm b}}L_{\rm b}^{-1}\frac{{\bf h_{\rm b}}\times{\bf e}_{\rm b}}{(1-e_{\rm b}^2)^{3/2}}
\\
&+\frac{3}{2}n_{\rm b}\frac{ M_\bullet}{M_{\rm b}}\frac{a_{\rm b}^3}{a_{\rm out}^3}L_{\rm b}^{-1}
\Big[5({\bf e}_{\rm b}\cdot\hat{\bf r}_{\rm out}){\bf h_{\rm b}}\times\hat{\bf r}_{\rm out}
\\
&~~~~~~~~~~~~~~~~
-2{\bf h_{\rm b}}\times{\bf e}_{\rm b}
-({\bf h}_{\rm b}\cdot\hat{\bf r}_{\rm out}){\bf e_{\rm b}}\times\hat{\bf r}_{\rm out} \Big]
\end{split}
\\
\begin{split}
\frac{\D{\bf h}_{\rm b}}{\D t}\bigg|_{\rm c}&=
\frac{3}{2}n_{\rm b}
\frac{ M_\bullet}{M_{\rm b}}\frac{a_{\rm b}^3}{a_{\rm out}^3}
\big[ 5L_{\rm b}({\bf e}_{\rm b}\cdot\hat{\bf r}_{\rm out})({\bf e}_{\rm b}\times\hat{\bf r}_{\rm out})
\\
&~~~~~~~~~~~~~~~~~
-L_{\rm b}^{-1}({\bf h}_{\rm b}\cdot\hat{\bf r}_{\rm out})({\bf h}_{\rm b}\times\hat{\bf r}_{\rm out})
\big]
\end{split}
\end{align}
\end{subequations}
where the subscript `c' denotes `conservative' \citep[see also][]{liu18}. Equations~(\ref{eq:eom_milankovitch}) preserve the binary's semi-major axis, and
consequently,
\begin{equation}\label{eq:hardening}
\frac{\dot{a}_{\rm b}}{a_{\rm b}}
=
\frac{2}{1-e_{\rm b}^2}\left[\frac{{\bf h}_{\rm b}\cdot\dot{\bf h}_{\rm b}}{L_{\rm b}^2}
+{\bf e}_{\rm b}\cdot\dot{\bf e}_{\rm b}\right]
\end{equation}
evaluates identically to zero in the absence of dissipation.

The energy and angular momentum losses due to GW emission are \citep{pete64}
\begin{subequations}\label{eq:gwr_milankovitch}
\begin{align}
\frac{\D{\bf e}_{\rm b}}{\D t}\bigg|_{\rm GWR}\!\!\!=&
-n_{\rm b} \frac{304}{15}\frac{\mu_{\rm b}}{M_{\rm b}}\left(\frac{r_g}{a_{\rm b}}\right)^{5/2}
\frac{(1+\tfrac{121}{304}e_{\rm b}^2)}{(1-e_{\rm b}^2)^{5/2}}{\bf e}_{\rm b}
\\
\frac{\D{\bf h}_{\rm b}}{\D t}\bigg|_{\rm GWR}\!\!\!=&
-n_{\rm b} \frac{32}{5}\frac{\mu_{\rm b}}{M_{\rm b}}\left(\frac{r_g}{a_{\rm b}}\right)^{5/2}
\frac{(1+\tfrac{7}{8}e_{\rm b}^2)}{(1-e_{\rm b}^2)^{5/2}}{\bf h}_{\rm b}
~,
\end{align}
\end{subequations}
or, from Equation~(\ref{eq:hardening}), one may obtain the more familiar expression \citep{pete64}
\begin{equation}\label{eq:hardening2}
\frac{\dot{a}_{\rm b}}{a_{\rm b}}\bigg|_{\rm GWR}
=
-\frac{1}{t_{\rm hard,0}}\frac{1}{4}\left(\frac{a_{{\rm b},0}}{a_{\rm b}}\right)^4
F(e_{\rm b})
\end{equation}
with $F(e_{\rm b})={\left(
1+\tfrac{73}{24}e_{\rm b}^2+\tfrac{37}{96}e_{\rm b}^4
\right)}{(1-e_{\rm b}^2)^{-7/2}}$~
and where 
\begin{equation}\label{eq:thard}
t_{\rm hard,0}=\frac{5 }{256}\frac{M_{\rm b}}{\mu_{\rm b}}\frac{a_{\rm b,0}^{3/2}}{\sqrt{{\cal G}M_{\rm b}}}
\left(\frac{r_g}{a_{\rm b,0}}\right)^{-5/2}
\end{equation}
is the initial hardening time, with $a_{\rm b,0}$ being
the initial semi-major axis.

In principle, the trajectory of the outer orbit is determined from
$\dot{\bf r}_{\rm out}=-{{\cal G}M_\bullet}({\bf r}_{\rm out}/|{\bf r}_{\rm out}|^3)+{\bf f}_{\rm extra}$ where
the additional force ${\bf f}_{\rm extra}$ is responsible for  migration within the AGN disk.
But we choose
instead to prescribe this $\dot{\bf r}_{\rm out}$ as a circular, zero-inclination orbit (i.e., the disk symmetry axis $\hat{\bf z}$ and the angular momentum orientation of the outer orbit $\hat{\bf h}_{\rm out}$ are aligned; Figure~\ref{fig:orbits}) Thus, in Equation~(\ref{eq:eom_milankovitch}),
we replace
\begin{equation}
{\bf r}_{\rm out}=a_{\rm out}(t)
\begin{pmatrix}
\cos \lambda_{\rm out}(t)\\
\sin \lambda_{\rm out}(t)\\
0
\end{pmatrix}~,
\end{equation}
where $a_{\rm out}(t)$ is a time-varying semi-major axis, shrinking at a {\it prescribed} migration rate
\begin{equation}\label{eq:migration_rate}
t_{\rm mig}\equiv-\frac{a_{\rm out}}{\dot{a}_{\rm out}}~,
\end{equation}
$\lambda_{\rm out}={\cal M}_{\rm out}+ \varpi_{\rm out}$ is the mean longitude,
$\varpi_{\rm out}=\omega_{\rm out}+\Omega_{\rm out}$ is the longitude of pericenter, and
\begin{equation}
 {\cal M}_{\rm out}= {\cal M}_{\rm out,0}
 +\int_0^t n_{\rm out}(t')dt'
\end{equation}
is the mean anomaly, where the outer orbit's mean motion
is $n_{\rm out}(t)=\sqrt{{\cal G}M_\bullet/a^3_{\rm out}(t)}$.

The full set of equations of motion is thus
\begin{subequations}\label{eq:milankovitch_eom}
\begin{align}
&\frac{\D{\bf e}_{\rm b}}{\D t}=\frac{\D{\bf e}_{\rm b}}{\D t}\bigg|_{\rm c}+
\frac{\D{\bf e}_{\rm b}}{\D t}\bigg|_{\rm GWR}~,
\\
&\frac{\D{\bf h}_{\rm b}}{\D t}=
\frac{\D{\bf h}_{\rm b}}{\D t}\bigg|_{\rm c}
+\frac{\D{\bf h}_{\rm b}}{\D t}\bigg|_{\rm GWR}~,
\\
&\frac{\D{r}_{{\rm out},x}}{\D t}= -\frac{a_{\rm out}}{t_{\rm mig}}\cos\lambda_{\rm out}
-a_{\rm out}\sin\lambda_{\rm out} n_{\rm out}~,
\\
&\frac{\D{r}_{{\rm out},y}}{\D t}=  -\frac{a_{\rm out}}{t_{\rm mig}}\sin\lambda_{\rm out}
+a_{\rm out}\cos\lambda_{\rm out} n_{\rm out}~.
\end{align}
\end{subequations}
\paragraph{Fiducial case}
We integrate Equations~(\ref{eq:milankovitch_eom}) numerically for a
a coplanar triple black hole system consisting
of a quasi-circular BHB ($e_{\rm b}=0.005$) with $M_{\rm b}=510M_\odot$, $q_{\rm b}=0.02$
orbiting a SMBH with $M_\bullet=10^8M_{\odot}$ and initial orbital distance of 900~au. For the migration of the BHB, we
prescribe $t_{\rm mig}$ in Equations~(\ref{eq:milankovitch_eom}) in terms of the outer orbital period $P_{\rm out}=2\pi/n_{\rm out}$
\begin{equation}\label{eq:beta_mig}
t_{\rm mig}=\beta_{\rm mig} P_{\rm out}
\end{equation}
where $\beta_{\rm mig}$ is a constant.

We carry out three examples with $\beta_{\rm mig}=$ 
$2\times10^4$, $3\times10^4$ and $4\times10^4$ (the first two are `fast migrators' and the third one is a `slow migrator') and present the results of these integrations in Figure~\ref{fig:eccentricity}. In the top panels,  we show the evolution of the outer orbit, represented by $n_{\rm out}$ (orange curves), which we compare to the evolution of the apsidal precession rate due to general relativistic effects
\begin{equation}\label{eq:apsidal_rate}
 \dot{\omega}_{\rm GR}=3n_{\rm b}\frac{r_g}{a_{\rm b}}\frac{1}{1-e_{\rm b}^2}~,
 \end{equation}
(blue curves). All three cases depicted in the Figure start with 
$\dot{\omega}_{\rm GR}>n_{\rm out}$. Subsequently, $n_{\rm out}$ and $\dot{\omega}_{\rm GR}$
increase in time due to BHB migration and hardening, respectively, but since
 $n_{\rm out}$ grows at a faster rate, commensurability crossing is possible. Note however, that while this crossing
 takes place in all three examples, it is only the third one (the `slow migration' case) that exhibits non-trivial
 behavior after $n_{\rm out}$ catches up to $\dot{\omega}_{\rm GR}$: both these frequencies start evolving in lockstep, as it occurs in cases of resonance capture.
 
Outside the resonant regime, the evolution of $\dot{\omega}_{\rm GR}$ is straightforward as long as the binary remains quasi-circular, in which case 
$\dot{\omega}_{\rm GR}\propto a_{\rm b,circ}^{-5/2}$, where
\begin{equation}\label{eq:ab_circ}
a_{\rm b,circ}(t)=a_{{\rm b},0}\left[1-\frac{t}{t_{\rm hard,0}}\right]^{1/4}
\end{equation}
is the well-known Peter's solution to Equation~(\ref{eq:hardening2}) 
for an initial binary separation of $a_{\rm b,0}$
when $e_{\rm b}\approx0$.  Similarly, 
the outer orbital frequency $n_{\rm out}\propto a_{\rm out}^{-3/2}$ attains a simple form because
our funcional choice for
$t_{\rm mig}$ (Equation~\ref{eq:beta_mig}) allows us to obtain $a_{\rm out}$ analytically as well:
\begin{equation}\label{eq:aout_mig}
a_{\rm out}(t)=a_{{\rm out},0}\left[1-\frac{3}{2}\frac{t}{t_{\rm mig,0}}\right]^{2/3}~,
\end{equation}
where $t_{\rm mig,0}\equiv \beta_{\rm mig}P_{\rm out,0}$
and $a_{{\rm out},0}$ is the initial condition.
These explicit dependencies on $t$ will allow us to solve for the crossing time $t_\star$ in advance of the numerical integration.
For the three examples shown in  Figure~\ref{fig:eccentricity}, commensurability crossing occurs at $t_\star=1.771\times10^4$~yr (left),
$t_\star=2.876\times10^4$~yr (middle),
 $t_\star=4.183\times10^4$~yr (right). 

At the time of crossing, there is a change in eccentricity evolution (middle panel of each column). This change is a moderate (factor of $\sim4$), one-time-only jump in the first and second examples (the `fast migrators'), but it appears to grow indefinitely for the `slow migrator' case, increasing 500-fold over a small
fraction of the migration timescale ($t_{\rm mig,0}=1.08\times10^6$~yr). As we describe in Section~\ref{sec:evection} below, this dramatic eccentricity
growth results from the system being captured into a state of evection resonance.

Finally, the bottom panel of each example in Figure~\ref{fig:eccentricity}  depicts the angle $\lambda_{\rm out}-\varpi_{\rm b}$ subtended
by the vectors ${\bf r}_{\rm out}$ and ${\bf e}_{\rm b}$ (see Figure~\ref{fig:orbits}). In the `fast migrator' cases on the left, this angle 
circulates freely between 0 and $2 \pi$;  in the `slow migrator' case, however, this angle librates close to $\tfrac{3}{2}\pi$ after the commensurability
has been crossed. Once again, this behavior can be indicative of resonance capture.

%%%%%%%%%%%%%%%%%%%%%%%%%%%%%%%%
%%%%%%%%%%%%%%%%%%%%%%%%%%%%%%%%
\begin{figure*}
\includegraphics[width=1.04\textwidth]{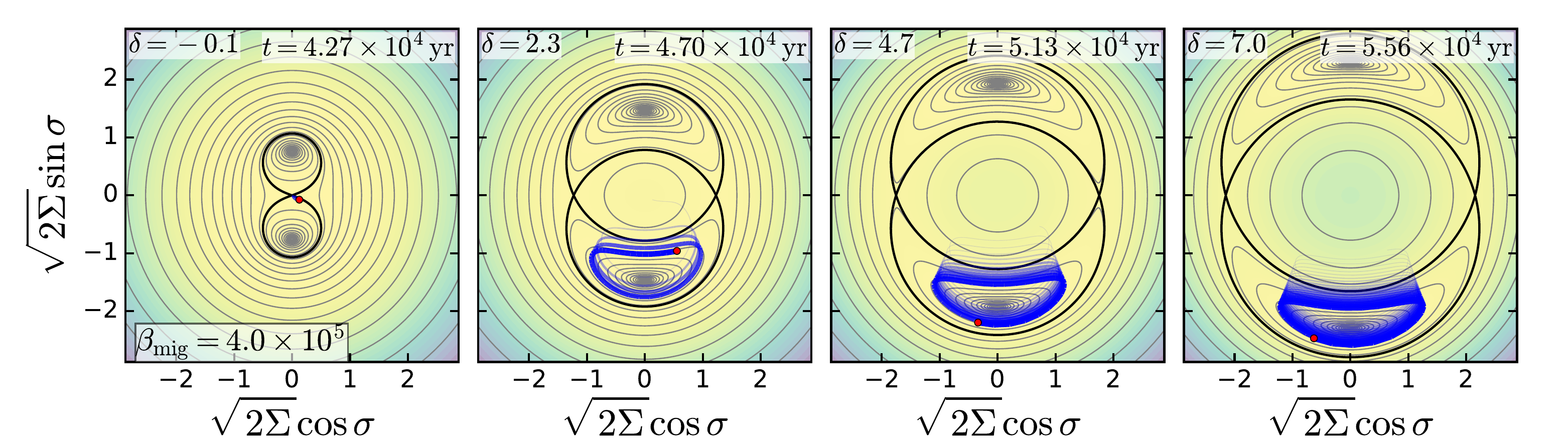}
\caption{Autonomous Hamiltonian ${\cal K}$ (Equation~\ref{eq:transformed_hamiltonian}) in Poincar\'e rectangular coordinates ($\eta\equiv\sqrt{2\Sigma}\sin\sigma$ 
 $\xi\equiv\sqrt{2\Sigma}\cos\sigma$), 
 evaluated with  $M_\bullet=10^8M_{\odot}$, $M_{\rm b}=510M_\odot$, $q_{\rm b}=0.02$ and for
$(a_{\rm b},a_{\rm out})$=
$(5.960\times10^{-2}, 864.1)$,
$(5.956\times10^{-2}, 860.4)$,
$(5.952\times10^{-2}, 856.7)$
and
$(5.947\times10^{-2},853)$ au (left to right panels). 
 Two bifurcations take place in the vicinity of $t_\star=4.183\times10^4$~yr, the time of commensurability crossing.
The blue trajectory depicts the numerical solution of the fiducial example (Figure~\ref{fig:eccentricity}),
written in Poincar\'e rectangular coordinates using the transformation $\sigma=\lambda_{\rm out}-\omega_{\rm b}-\Omega_{\rm b}$
and $\Sigma= L_{\rm b}(1-\sqrt{1-e_{\rm b}^2})$, with $e_{\rm b}=|{\bf e}_{\rm b}|$.
\label{fig:hamiltonian}}
\end{figure*}
%%%%%%%%%%%%%%%%%%%%%%%%%%%%%%%%
%%%%%%%%%%%%%%%%%%%%%%%%%%%%%%%%
%%%%%%%%%%%%%%%%%%%%%%%%%%%%%%%%%%%%%%%%%%%%%%%%%%%%%%%%%

 %%%%%%%%%%%%%%%%%%%%%%%%%%%%%%%%%%%%%%%%%%%%%%%%%%%%%%%%%
\subsection{Evection Resonance}\label{sec:evection}
\subsubsection{Overview of Nonlinear Resonance}
Nonlinear resonance capture is the mechanism by which the circulating trajectories of a Hamiltonian
${\cal K}(q,p;\delta)$ can transition into librating ones 
as a parameter $\delta$ of the system changes {\it slowly} in time \citep[see][]{nei75,tim78,hen82,car86,hen93}.
This process may involve the {\it appearance} of a separatrix, the {\it crossing} of a separatrix, or a combination thereof. Away from the separatrix, the evolution of the system is governed by the conservation
of the action $J=(2\pi)^{-1}\oint pdq$ or adiabatic invariant, provided that ${\cal K}$ changes slowly enough as to approximate it by a sequence
of ``frozen'' autonomous Hamiltonians \citep[e.g.,][]{landau1969,jose1998}. This adiabaticity allows the resonant/librating trajectories to be drifted along with the fixed points at the center of the libration region, which change as a function of $\delta$. This drift can lead to an arbitrary growth  of the canonical momenta $p$ while conserving $J$. In the example at hand, we are interested in the canonical momentum  
$\Gamma=\mu_{\rm b}\sqrt{{\cal G} M_{\rm b}}(1-\sqrt{1-e_{\rm b}^2})$, whose resonant growth is equivalent to growth in orbital eccentricity.

\subsubsection{Resonant Hamiltonian}
We can identify 
the relevant resonant terms in the single-average Hamiltonian~(\ref{eq:averaged_hamiltonian}).
Assuming that the hierarchical triple is coplanar and
using modified Delaunay canonical coordinates
$\lambda=l_{\rm b}+\omega_{\rm b}+\Omega_{\rm b}$,
$\gamma=-\omega_{\rm b}-\Omega_{\rm b}$, and momenta
$\Lambda=L_{\rm b}~$,
$\Gamma= L_{\rm b}(1-\sqrt{1-e_{\rm b}^2})$
\citep[e.g.,][]{mur00,morbidelli2002},
we obtain
\begin{equation}\label{eq:poincare_hamiltonian}
\begin{split}
{\cal H}' =&-n_{\rm b}\frac{3r_g}{a_{\rm b}}
\Lambda\left(1-\frac{\Gamma}{\Lambda}\right)^{-1}
\\
&-\frac{n_{\rm out}^2}{n_{\rm b}}\frac{\Lambda}{4}
\bigg[1+\frac{3}{2}\frac{\Gamma}{\Lambda}\left(2-\frac{\Gamma}{\Lambda}\right)\\
&~~~~~~~~~~~~~~~
+\frac{15}{2}\frac{\Gamma}{\Lambda}\left(2-\frac{\Gamma}{\Lambda}\right)\cos2\left(\lambda_{\rm out}+\gamma\right)
\bigg].\!\!\!\!\!\!\!\!
\end{split}
\end{equation}
To remove this explicit time dependence of ${\cal H}'$, we introduce the angle
\begin{equation}\label{eq:resonant_angle}
\sigma\equiv \lambda_{\rm out}+\gamma=\lambda_{\rm out}-\omega_{\rm b}-\Omega_{\rm b}
\end{equation}
and replace the canonical pair $(\gamma,\Gamma)$ with a new
one $(\sigma,\Sigma)$ via a time-dependent Type-2 generating function $F_2=\lambda\Lambda'+(\lambda_{\rm out}(t)+\gamma)\Sigma$
\citep{tou98}. The new momenta are $\Lambda'=\Lambda$ and $\Sigma=\Gamma$, and the transformed Hamiltonian is, after dropping constants and
unnecessary primes, 
\begin{equation}\label{eq:transformed_hamiltonian}
\begin{split}
{\cal K} =&n_{\rm out}\Sigma-n_{\rm b}\frac{3r_g}{a_{\rm b}}
\Lambda\left(1-\frac{\Sigma}{\Lambda}\right)^{-1}
\\
&-\frac{3}{8}\frac{n_{\rm out}^2}{n_{\rm b}}
\Sigma\left(2-\frac{\Sigma}{\Lambda}\right)\left(1+5\cos2\sigma
\right),\!\!\!\!\!\!
\end{split}
\end{equation}
which is autonomous, and of one degree of freedom, and therefore describes and integrable system.

%%%%%%%%%%%%%%%%%%%%%%%%%%%%%%%%
%%%%%%%%%%%%%%%%%%%%%%%%%%%%%%%%
\begin{figure}
\includegraphics[width=0.47\textwidth]{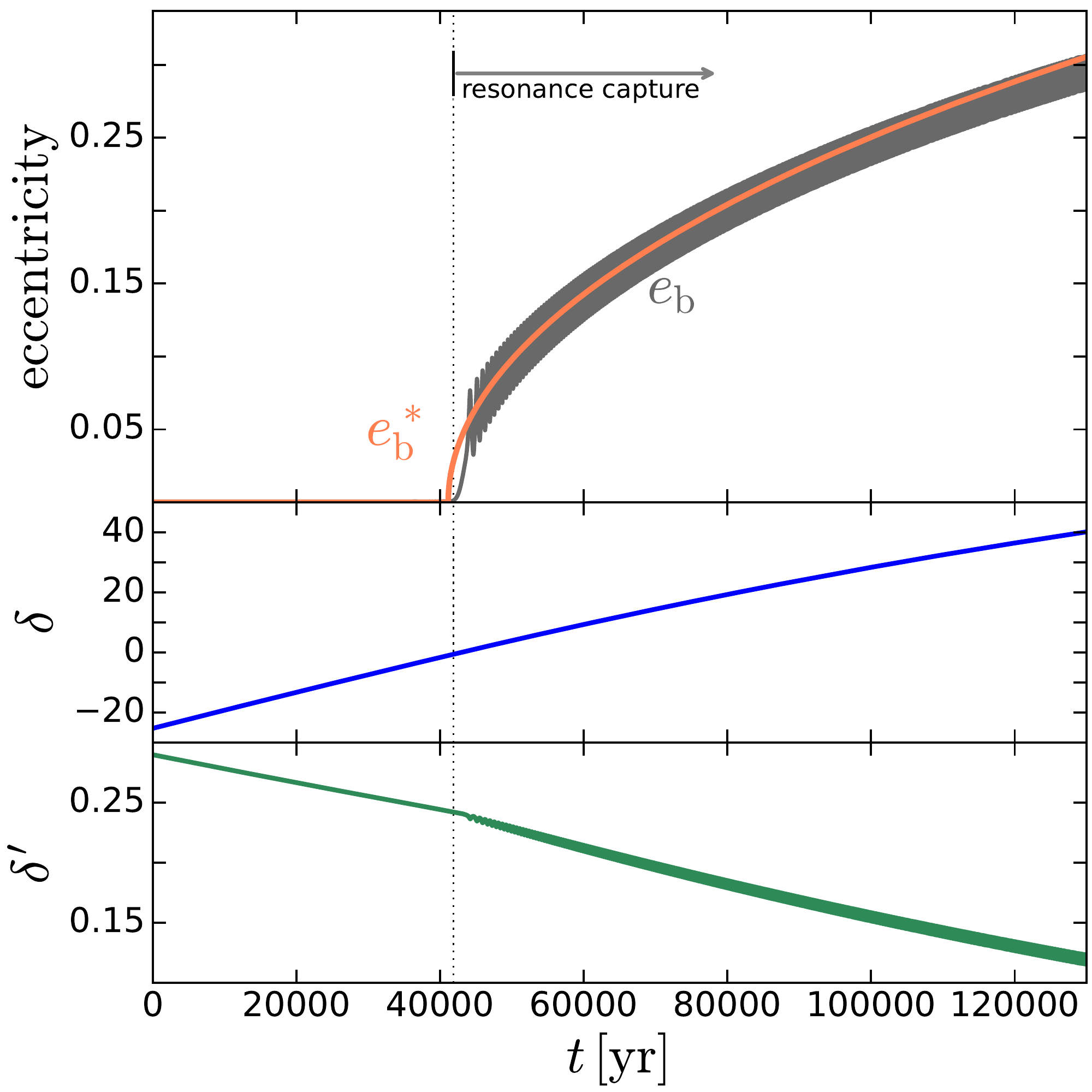}
\caption{Top panel:  fiducial example of Figure~\ref{fig:eccentricity} (right panel), this time comparing the numerically obtained $e_{\rm b}$ (gray) to
the analytic solution $e_{\rm b}^\ast$ of Equation~\ref{eq:ecc_fixed_point} (orange). Middle panel:  evolution of the drift parameter $\delta$ (Equation~\ref{eq:delta})
as a function of time.  Bottom panel:  evolution of the dimensionless drift rate $\delta'$ (Equation~\ref{eq:delta_prime_2}), confirming that $\delta'\lesssim0.25$
at the time of commensurability crossing \citep{qui06}. The vertical dotted line depicts $t_\star$, the time of commensurability crossing.
\label{fig:capture}}
\end{figure}
%%%%%%%%%%%%%%%%%%%%%%%%%%%%%%%%
%%%%%%%%%%%%%%%%%%%%%%%%%%%%%%%%
%%%%%%%%%%%%%%%%%%%%%%%%%%%%%%%%%%%%%%%%%%%%%%%%%%%%%%%%%

The topology of ${\cal K}$ is largely equivalent to that of the ``second fundamental model of resonance'' of order $k=2$  (\citealp{hen83,bor84,mal90}; see Appendix~\ref{app:fundamental}). In Figure~\ref{fig:hamiltonian}, we highlight this topology by depicting ${\cal K}$ using the values of $a_{\rm b}(t)$ and $a_{\rm out}(t)$ evaluated at different times, according to the fiducial
model with $\beta_{\rm mig}=4\times10^5$. In the figure, we also include the numerical solution from Figure~\ref{fig:eccentricity} in $\xi,\eta$ coordinates (blue trajectories).
This numerical trajectory starts off circulating around the origin ($\sigma$ takes values from 0 to $2\pi$),
but is subsequently displaced, keeping up with the fixed point, which  gradually shifts toward higher values of $\Sigma$ following two bifurcations. As this displacement occurs, $\sigma$ (Equation~\ref{eq:resonant_angle}) transitions from circulating form 0 to $2\pi$ to librating around $\tfrac{3}{2}\pi$.

When $\Sigma/\Lambda\ll1$, the fixed point  is given by \citep{bor84}:
\begin{equation}\label{eq:fixed_points}
\sigma^\ast=\pm \frac{\pi}{2}~,\;\;\;\;
\Sigma^*= \frac{5}{2}\frac{n_{\rm out}^2}{n_{\rm b}^2}
\frac{a_{\rm b}}{r_g}\Lambda \frac{1+\delta}{2},
\;\;\;\;\text{if}\;\;\;\delta>-1
\end{equation}
where the quantity
\begin{equation}\label{eq:delta}
\delta\equiv-\frac{3}{5}+\frac{2n_{\rm b}}{15n_{\rm out}}\left(1-3\frac{n_{\rm b}}{n_{\rm out}}\frac{r_g}{a_{\rm b}}\right)
\end{equation}
(see Appendix~\ref{app:fundamental})  is  the single free parameter that appears in the second fundamental model of \citet{hen83}. Then, after reinstating our original coordinates, the fixed point (\ref{eq:fixed_points}) can be written in terms of the binary eccentricity as
\begin{equation}\label{eq:ecc_fixed_point}
e_{\rm b}^\ast \approx\left(\frac{5}{2}\frac{n_{\rm out}^2}{n_{\rm b}^2}
\frac{a_{\rm b}}{r_g}\right)^{1/2}\sqrt{1+\delta}~.
\end{equation}
Thus, growth in $\delta$ implies growth in eccentricity. Note, however, that  $\delta$ may well decrease in time. 
Indeed, as some compact binaries can harden faster than they migrate, the term in parenthesis in Equation~(\ref{eq:delta}) 
may be always negative, which may shift $\delta$ inexorably toward more negative values. 

The equilibrium eccentricity $e_{\rm b}^\ast$ well captures the behavior of the numerical solution, as depicted in the top panel of Figure~\ref{fig:capture}. The figure also shows the evolution of $\delta$ (middle panel),  which grows from large negative values to large positive values, attaining a value of $-3/5$ when the commensurability (\ref{eq:commensurability}) is crossed. Finally, the bottom panel shows the ``drift rate'' $\delta'$, which is the rescaled time derivative of $\delta$ (see Section~\ref{sec:conditions} below). Consistent with the results of \citet{qui06}, which state that resonance capture occurs when $\delta'(t_\star)\approx0.25$ (see below).

\subsubsection{Likelihood of Resonance Capture}\label{sec:conditions}
The dramatically different outcomes of Figure~\ref{fig:eccentricity}
underscore the importance of the migration rate in determining whether capture into an evection resonance
can be guaranteed.

%%%%%%%%%%%%%%%%%%%%%%%%%%%%%%%%
%%%%%%%%%%%%%%%%%%%%%%%%%%%%%%%%
\begin{figure*}
\includegraphics[width=0.47\textwidth]{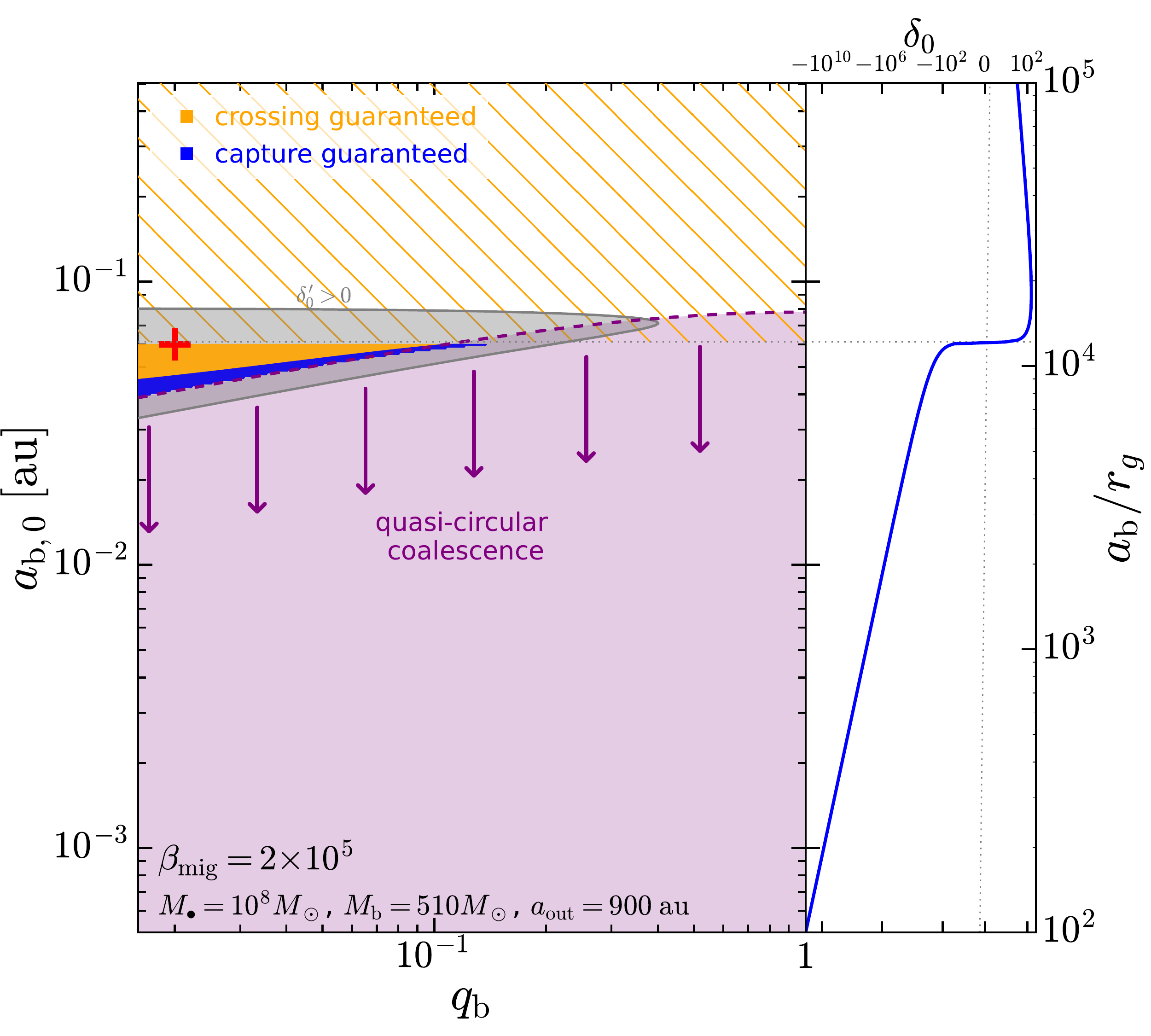}
\hspace{0.2in}
\includegraphics[width=0.47\textwidth]{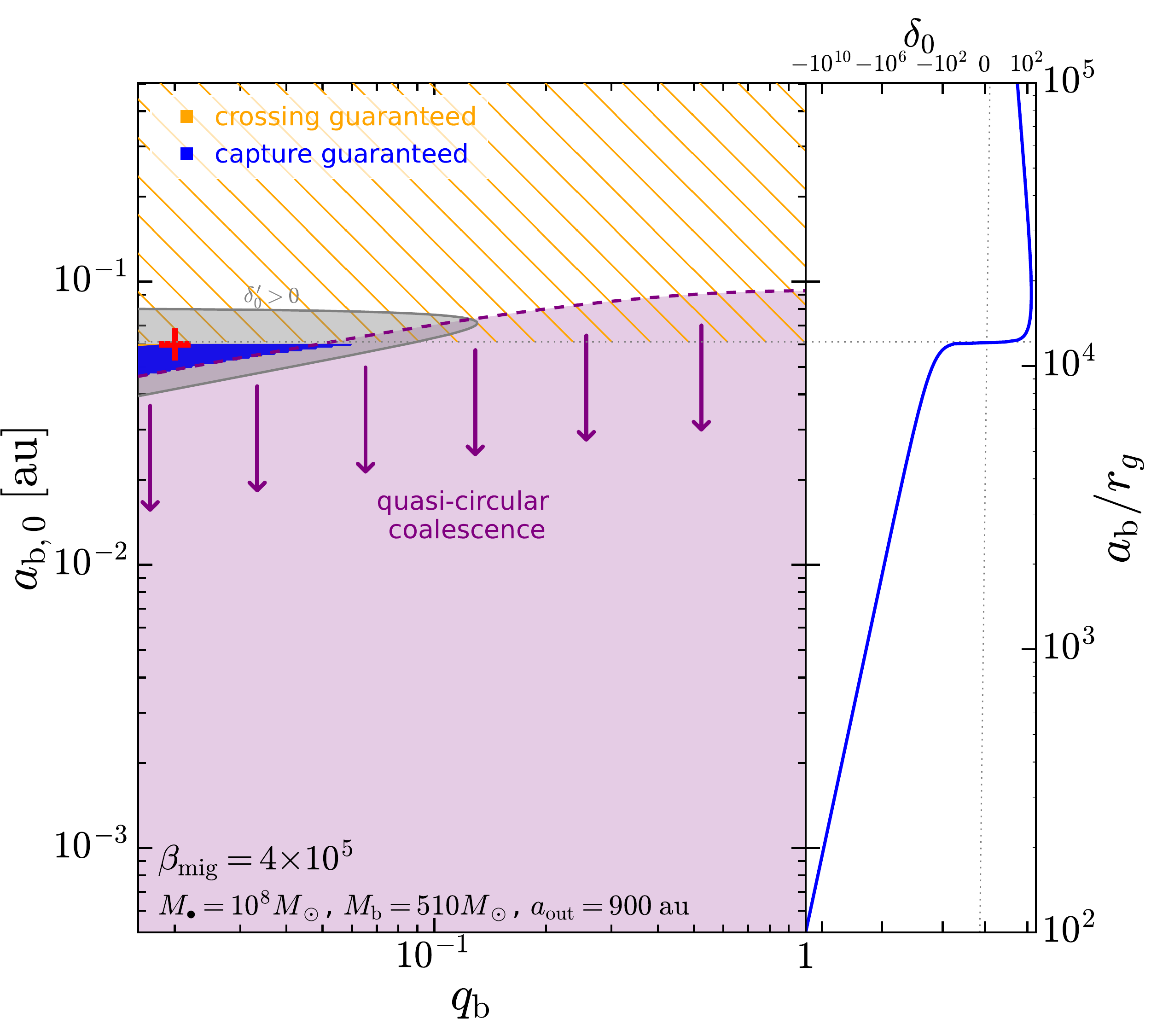}
\caption{The likelihood of evection resonance capture visualized in the
parameter space of inner mass ratio ($q_{\rm b}$) and initial semimajor axis ($a_{\rm b,0}$)
for given values of $M_\bullet$, $M_{\rm b}$, $a_{\rm out,0}$ and $\beta_{\rm mig}$. The parameter is subdivided into three regions, which may overlap:
(1) purple, (2) orange (solid and hatched) and (3) blue. Purple depicts a population of BHBs that will merge in a quasi-circular fashion (Equation~\ref{eq:ab_circ})
before migrating through the disk  (i.e., $t_{\rm hard,0}<\tfrac{2}{3} t_{\rm mig,0}$); except for tidal disruptions due to the SMBH, these binaries will merge as if they existed in vacuum.
Orange depicts the set of parameters/initial conditions for which the condition $\delta=-\tfrac{3}{5}$ (i.e., $\dot{\omega}_{\rm GR}=n_{\rm out}$) will be eventually met or ``crossed"; hatched orange represents 
crossing with $\delta'_\star<0$ ($\delta$ decreasing) and solid orange represents crossing with $\delta'_\star\geq0$ ($\delta$ increasing). Blue represents `guaranteed capture' according to all criteria of Section~\ref{sec:conditions} and is always contained within the solid orange region. In addition, we highlight the area for which the initial conditions satisfy $\delta'_0>0$ (gray) and we include an ancillary panel showing the initial value of the drift parameter $\delta_0$ as a function of $a_{\rm b,0}$ to emphasize the wide range of values
$\delta$ may take. Left panel: likelihood of evection capture for the parameters
$M_\bullet=10^8M_\odot$, $M_{\rm b}=510M_\odot$, $a_{\rm out,0}=900$~au and $\beta_{\rm mig}=2\times10^5$,
(as in the left panel of Figure~\ref{fig:eccentricity}. 
 The initial conditions  ($q_{\rm b},a_{\rm b,0}$)=(0.02,0.06~au) of the fiducial example (Section~\ref{sec:equations}) are
depicted as a red cross, which falls in the orange region, indicating that crossing takes place, but capture is not guaranteed, as already evidenced by the left panel of Figure~\ref{fig:eccentricity}.
Right panel: same as on the left, but for $\beta_{\rm mig}=4\times10^5$. In this case, the orange and blue region nearly overlap, with the red cross
indicating that `resonance capture' is guaranteed, as was already seen from the right panel of Figure~\ref{fig:eccentricity}.  In both panels, the likelihood of evection capture (blue region) grows as $q_{\rm b}$ get smaller.
\label{fig:parameters}}
\end{figure*}
%%%%%%%%%%%%%%%%%%%%%%%%%%%%%%%%
%%%%%%%%%%%%%%%%%%%%%%%%%%%%%%%%

To guarantee capture into resonance under the ``second fundamental model'', three requirements must be met \citep[e.g.][]{qui06}:
\begin{itemize}
     \item[(i)] Passage through the commensurability must be {\it slow}  \citep{hen82,hen93}.
     This requirement is tantamount to the adiabatic theorem, which typically states that, in order to preserve adiabaticity, the drift  parameter $\delta$ must 
     satisfy $\dot{\delta} \omega_0\ll1$, where $\omega_0$ is the small oscillations frequency at the fixed point \citep{landau1969}. In practice, however,
     there is a sharp transition between the `too slow' and the `too fast' regimes. \citet{qui06} has numerically concluded that if
      \begin{equation}\label{eq:delta_prime}
\delta'\equiv    \frac{4}{15}\frac{n_{\rm b}}{n_{\rm out}^2}\,\frac{{\rm d}\delta}{{\rm d}t}
\lesssim 0.25
      \end{equation}
when $\delta=0$,
      then the probability of capture is almost certain  (see also \citealp{fri01})\footnote{%%%%%%%%%%%%%%%%%%%%%
Incidentally, we have numerically confirmed that the certainty of capture is ``fuzzier'' for the $k=2$ Henrard-Lemaitre Hamiltonian
than for $k=1$ (figures 2 and 3 of \citealp{qui06}). This effect is a consequence of the double bifurcation undergone by $\hat{\cal K}$
at $\delta=-1$ and $\delta =0$ (Appendix~\ref{app:fundamental}). Consequently, there is a small probability that a trajectory crosses a separatrix once to be captured
temporarily into resonance, only to cross a second separatrix, and end up in the inner  circulating (non-resonant) region.
}.
      
    \item[(ii)] The commensurability
    must be crossed in a specific direction: from $n_{\rm out}<\dot{\omega}_{\rm GR}$ toward
    $n_{\rm out}>\dot{\omega}_{\rm GR}$. In other words,  $\dot{\delta}>0$ when $\delta=-\tfrac{3}{5}$.

        \item[(iii)] The initial action $(2\pi)^{-1}\oint \Sigma d\sigma$  must be smaller than the area enclosed by separatrix at the time of bifurcation \citep{hen82}. If this requirement is not satisfied, capture is probabilistic (\citealp{hen82,hen83,bor84}; see also \citealp{yod79a}). This requirement translates on a maximum initial eccentricity $e_{\rm b,0}$ that makes capture certain:
\begin{equation}
e_{\rm b,0}\leq\sqrt{\frac{5}{2}\frac{a_{\rm b}}{r_g}}\left(\frac{n_{\rm out}}{n_{\rm b}}\right)
\end{equation}
\end{itemize}

With the exception of condition (iii), these requirements can be quite severe, limiting the parameter space that can successfully produce eccentricity growth from an evection resonance.
But above all, the drift rate $\delta'$ is the main hurdle to producing numerous evection resonance captures in AGN disks. 

We can write $\delta'$ as
\begin{equation}\label{eq:delta_prime_2}
\begin{split}
\delta'=\frac{4}{75}\frac{n_{\rm b}^2}{n_{\rm out}^3}
&\bigg[-\frac{\dot{a}_{\rm b}}{a_{\rm b}}\left(1-8\frac{n_{\rm b}}{n_{\rm out}}\frac{r_g}{a_{\rm b}}\right)
\\
&
+\frac{\dot{a}_{\rm out}}{a_{\rm out}}\left(1-6\frac{n_{\rm b}}{n_{\rm out}}\frac{r_g}{a_{\rm b}}\right)
\bigg]~.
\end{split}
\end{equation}
Equation~(\ref{eq:delta_prime_2})  clarifies the competing roles that binary migration and coalescing play in the onset of evection, and why the problem we are studying is quantitatively different from the lunar evection resonance. The expansion of the Moon's orbit implies that $\dot{\varpi}_{\rm b}$ decreases while $n_{\rm out}$ remains constant, allowing for the commensurability to be crossed in the right direction (condition (ii)). For a BHB, on the other hand, the very nature of orbital decay implies that $\dot{\varpi}_{\rm b}\approx \dot{\omega}_{\rm GR}$ is always increasing. Thus, to cross the commensurability in the right direction, one needs $n_{\rm out}$ to grow even faster than $ \dot{\omega}_{\rm GR}$ does (Figure~\ref{fig:eccentricity}, top left panel);  however if $n_{\rm out}$ grows too fast, then the commensurability can be crossed too quickly, hence violating condition (i) (Figure~\ref{fig:eccentricity}, top right panel). Conversely, if migration is too slow, the the commensurability might never be crossed at all.

%%%%%%%%%%%%%%%%%%%%%%%%%%%%%%%%
%%%%%%%%%%%%%%%%%%%%%%%%%%%%%%%%
\begin{figure}
\includegraphics[width=0.47\textwidth]{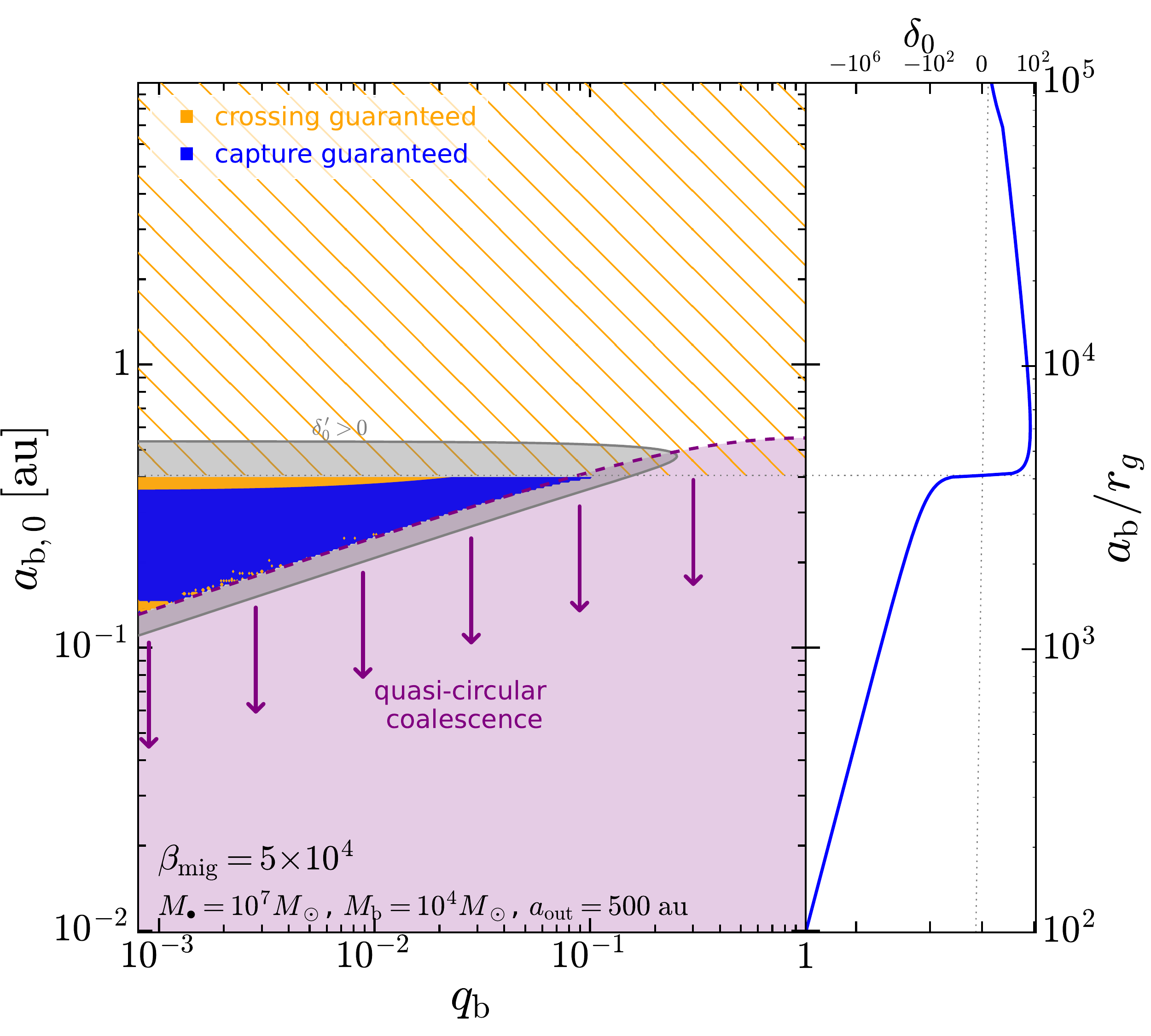}
\caption{Same as Figure~\ref{fig:parameters} but for parameters
$\beta_{\rm mig}=5\times10^4$, $M_\bullet=10^7M_\odot$, $M_{\rm b}=10^4M_\odot$ and
$a_{\rm out}=500$~au. For these parameters, the likelihood of resonance capture has been significantly increased.
As in Figure~\ref{fig:parameters} , the likelihood of evection capture (blue region) grows as $q_{\rm b}$ get smaller.
\label{fig:parameters2}}
\end{figure}
%%%%%%%%%%%%%%%%%%%%%%%%%%%%%%%%
%%%%%%%%%%%%%%%%%%%%%%%%%%%%%%%%

According to condition (ii) above,
the relevant quantity for resonance capture is 
$\delta'_\star\equiv \delta'(t_\star)$, i.e.,
\begin{equation}\label{eq:delta_prime_star}
\delta'_\star=\frac{2}{75\pi}\frac{a_{{\rm b},\star}^2}{9r_g^2}
\bigg[\beta_{\rm mig}^{-1} -\frac{5}{3}
\frac{2\pi}{t_{\rm hard,0}}\frac{a_{{\rm out},\star}^{3/2}}{\sqrt{{\cal G}M_\bullet}}\frac{1}{4}\left(\frac{a_{{\rm b},0}}{a_{{\rm b},\star}}\right)^4
\bigg]
\end{equation}
where $a_{{\rm b},\star}=a_{{\rm b}}(t_\star)$ and  $a_{{\rm out},\star}=a_{{\rm out}}(t_\star)$ are the semi-major axes evaluated at the time
of commensurability crossing. For BHBs that are initially quasi-circular, $t_{\star}$ is obtained from numerically solving
$n_{\rm out}=\dot{\omega}_{\rm GR}$ (Equation~\ref{eq:apsidal_rate}) when $a_{\rm b}$ and $a_{\rm out}$  are
given by Equations~(\ref{eq:ab_circ}) and~(\ref{eq:aout_mig}), respectively.
In summary, if given  $\beta_{\beta}$, $M_\bullet$, $M_{\rm b}$,
 $q_{\rm b}$, $a_{\rm out,0}$ and  $a_{\rm b,0}$, we can know in advance
 whether a system will cross the commensurability in the right direction, and whether this crossing will be place slow enough as to result in resonance capture.

%%%%%%%%%%%%%%%%%%%%%%%%%%%%%%%%
%%%%%%%%%%%%%%%%%%%%%%%%%%%%%%%%
\begin{figure}
\includegraphics[width=0.49\textwidth]{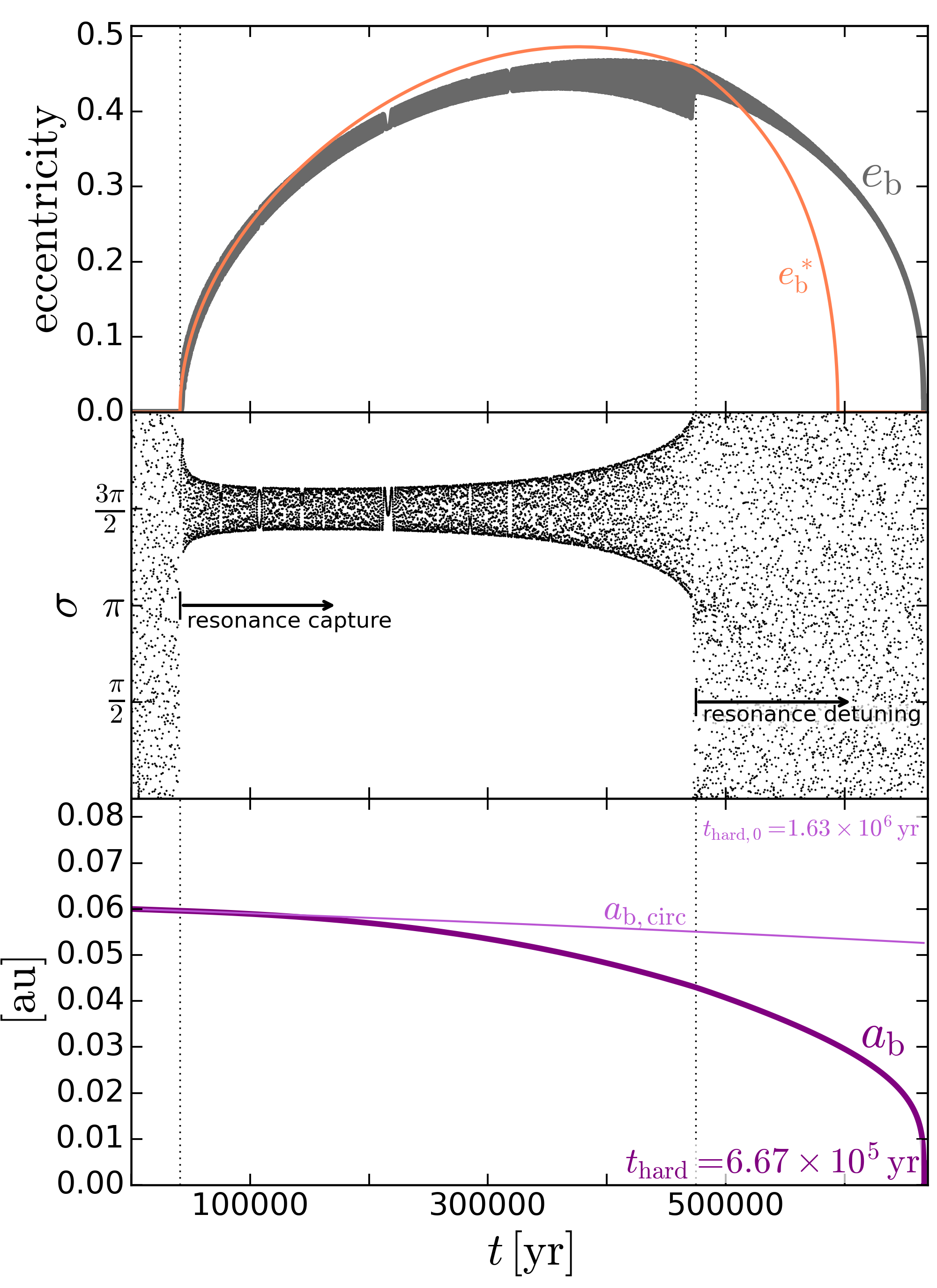}
\caption{Long-term behavior of the fiducial example from Figure~\ref{fig:eccentricity}'s right panel. Top panel: evolution of $e_{\rm b}$ (gray curve)
through resonance capture, eccentricity growth, resonance detuning and final circularization. The analytic solution $e_{\rm b}^\ast$ (Equation~
\ref{eq:ecc_fixed_point}, orange curve) closely tracks
$e_{\rm b}$ up to the moment of resonance detuning, after which the two curves start to diverge. Middle panel: the time evolution of the canonical angular coordinate 
$\sigma$ (Equation~\ref{eq:resonant_angle}, the ``resonant angle'') illustrates how the system is first captured into resonance ($\sigma$ transitions from circulation to libration)
and then leaves as the resonance is detuned  ($\sigma$ transitions  back to circulation). Bottom panel: evolution of $a_{\rm b}$ (thick purple line) throughout the resonance capture and detuning process. For comparison, we show $a_{\rm b, circ}$ (Equation~\ref{eq:ab_circ}, thin purple line), which describes the coalescence of a quasi-circular binary that does not experience resonance capture.  The evection resonance effectively accelerates the coalescence of this binary, shortening its merger time by a factor of $\simeq2.5$.
\label{fig:resonance}}
\end{figure}
%%%%%%%%%%%%%%%%%%%%%%%%%%%%%%%%
%%%%%%%%%%%%%%%%%%%%%%%%%%%%%%%%
%%%%%%%%%%%%%%%%%%%%%%%%%%%%%%%%%%%%%%%%%%%%%%%%%%%%%%%%%

In Figure~\ref{fig:parameters} we show the likelihood of resonance capture for the parameters
$M_\bullet=10^8M_{\odot}$, $M_{\rm b}=510M_\odot$,
$a_{\rm out,0}=900$~au
(e.g., Figure~\ref{fig:eccentricity}), and for range of values of
$a_{\rm b,0}$ and $q_{\rm b}$. The left panel depicts the case of $\beta_{\rm mig}=2\times10^5$ while 
the right panels shows $\beta_{\rm mig}=4\times10^5$.
In both panels, the bottom end of the figure (purple region) represents the binaries that are compact enough to coalesce quasi-circularly before reaching the center of the AGN disk ($t=\tfrac{2}{3}t_{\rm mig,0}$; Equation~\ref{eq:aout_mig}). Conversely, the upper region of each panel corresponds to binaries that are too wide to satisfy the $\dot{\omega}_{\rm GR}> n_{\rm out}$ condition at $t=0$ (violating condition (ii) above). In between these two regions, a narrow band of parameter space allows for commensurability crossing (solid orange, satisfying conditions (ii) and (iii) above). Within this region, a even smaller subset of parameters guarantees resonance capture (solid blue, satisfying conditions (i), (ii) and (iii) above).
The red cross in both panels represents the fiducial cases of
Figure~\ref{fig:eccentricity}: $q_{\rm b}=0.02$ and
$a_{\rm b,0}=0.06$~au. As expected the Figure~\ref{fig:eccentricity}, if $\beta_{\rm mig}=2\times10^5$, the cross falls within the `orange region' (crossing but no capture), but if  $\beta_{\rm mig}=4\times10^5$, the cross lies within the `blue region' (capture guaranteed).

Although Figure~\ref{fig:parameters} illustrates that, for the fiducial parameters explored thus far, evection resonance capture is infrequent (rates of $\lesssim1\%$),
Figure~\ref{fig:parameters2} paints a radically different picture. Choosing now $M_\bullet=10^7M_{\odot}$, $M_{\rm b}=10^4M_\odot$ and $a_{\rm out,0}=500$~au, we find
that the blue and orange regions of parameter space are now larger, and nearly exactly overlapping, meaning that crossing the commensurability alone nearly guarantees capture into resonance. In this case, 
amounting to a crossing/capture rates amount to $\sim10\%$, a significant increase from the previous example.

The increased in rates of Figure~\ref{fig:parameters2} obey primarily to more massive BHBs exhibiting faster apsidal precession, expanding the region of parameter space that satisfies $\dot{\omega}_{\rm GR}>n_{\rm out}$ (condition (ii) above). But more massive binaries coalesce more
quickly as well, and thus a decrease in $\beta_{\rm mig}$ allows $n_{\rm out}$ to catch up to $\dot{\omega}_{\rm GR}$, while also shortening
the amount of time binaries spend in the disk.

Another general feature of Figures~\ref{fig:parameters} and~\ref{fig:parameters2} is the increased crossing/capture rates for smaller values of $q_{\rm b}$. This is explained by the sensitivity of the merger timescale $t_{\rm hard,0}$ (Equation~\ref{eq:thard}) on $\mu_{\rm b}=q_{\rm b}M_{\rm b}/(1+q_{\rm b})$: equal-mass binaries coalesce too quickly for evection too operate. Thus, the direct coalescence regime overlaps with the $n_{\rm out}>\dot{\omega}_{\rm GR}$ regime, entirely forbidding capture.
%%%%%%%%%%%%%%%%%%%%%%%%%%%%%%%%
%%%%%%%%%%%%%%%%%%%%%%%%%%%%%%%%
\begin{figure*}
\centering
\includegraphics[width=0.30\textwidth]{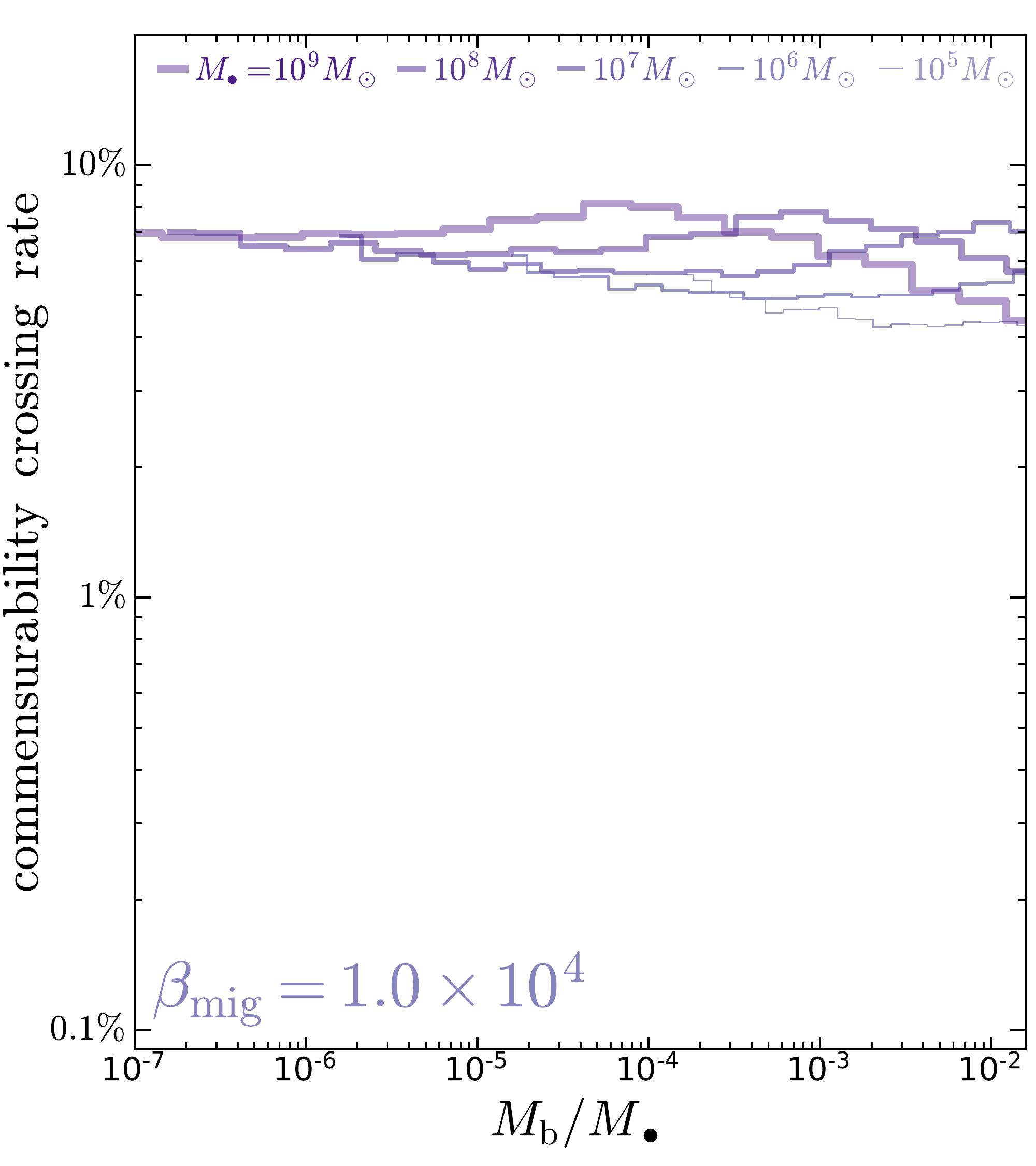}
\includegraphics[width=0.30\textwidth]{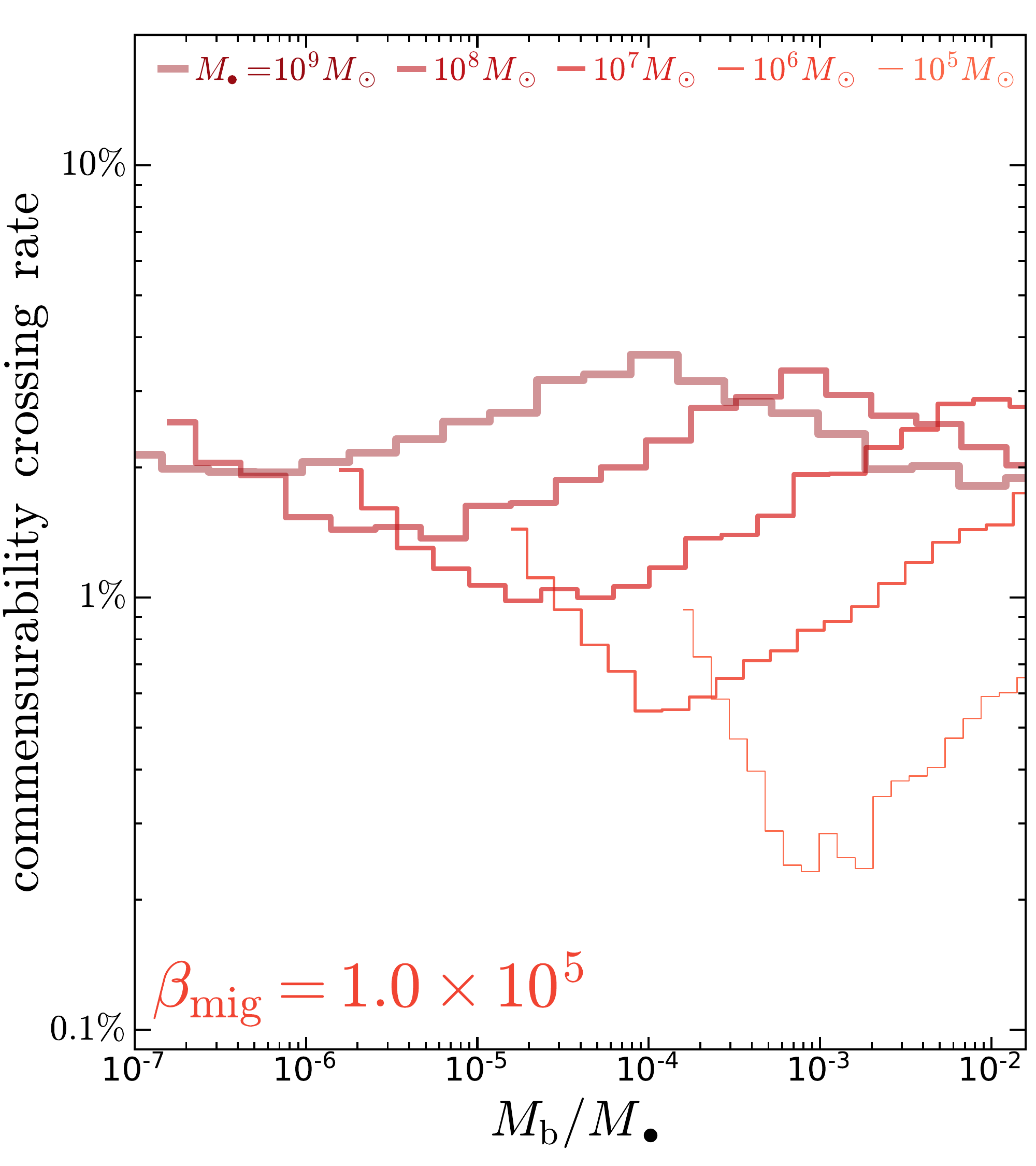}
\includegraphics[width=0.30\textwidth]{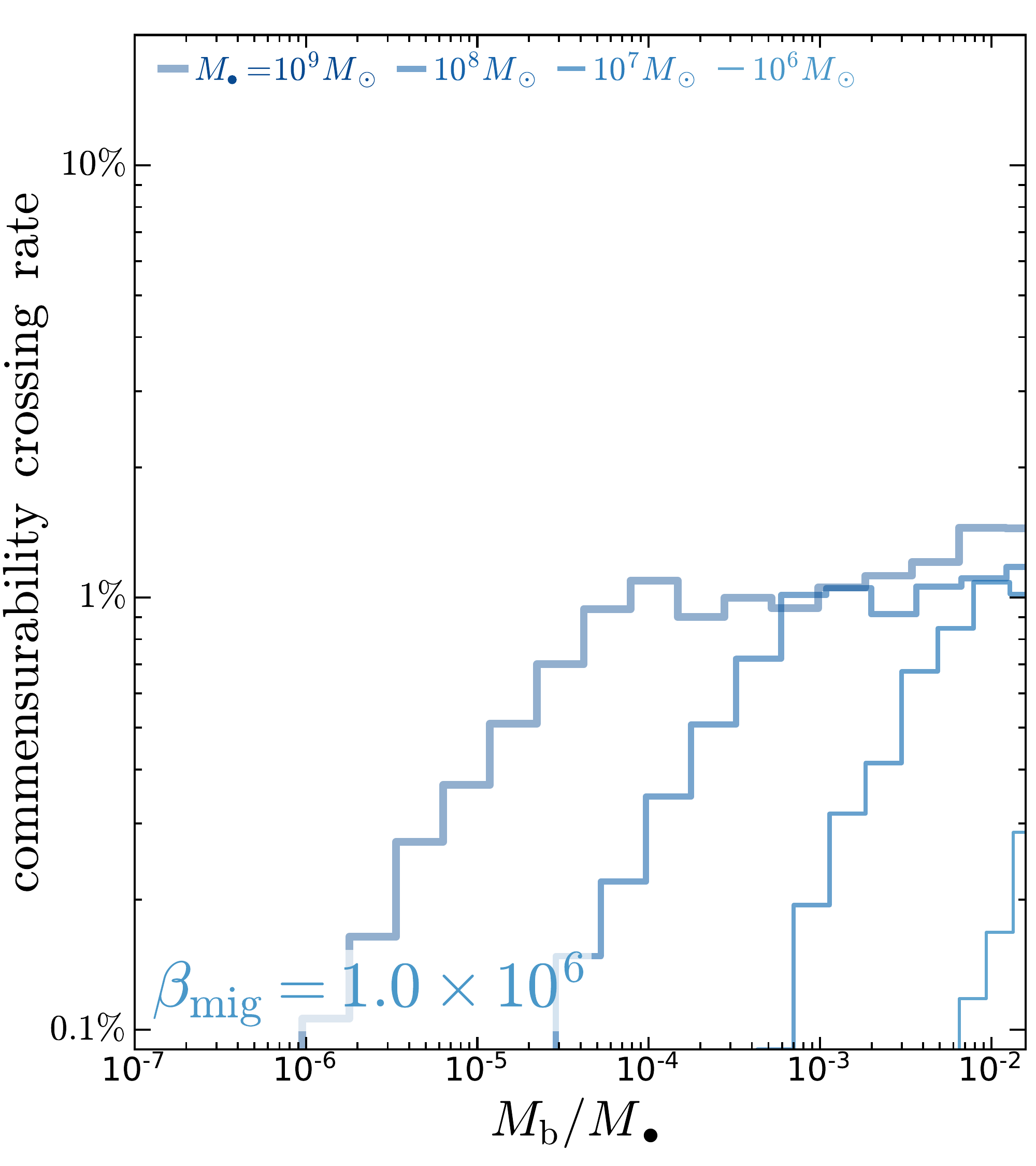}
\caption{Commensurability crossing rates. Commensurability is considered crossed if
$\delta_0<-\tfrac{3}{5}$ and  at any later point $\delta(t)>-\tfrac{3}{5}$ (i.e., $\dot{\omega}_{\rm GR}/n_{\rm out}$ goes from ${>}1$ to ${<}1$; see condition (ii) in
Section~\ref{sec:conditions} above).
Rates are obtained for numerous systems generated from a Monte Carlo sample of $q_{\rm b}$, $a_{\rm b}$ and $a_{\rm out}$, with $M_\bullet$, $M_{\rm b}$ and$\beta_{\rm mig}$ fixed. The systems with faster migration (smaller $\beta_{\rm mig}$) exhibit higher
rates of commensurability crossing, and are thus more likely to result in resonance captures and the resulting eccentric BHB mergers.
\label{fig:crossing_rate}}
\end{figure*}
%%%%%%%%%%%%%%%%%%%%%%%%%%%%%%%%
%%%%%%%%%%%%%%%%%%%%%%%%%%%%%%%%
%%%%%%%%%%%%%%%%%%%%%%%%%%%%%%%%%%%%%%%%%%%%%%%%%%%%%%%%%

%%%%%%%%%%%%%%%%%%%%%%%%%%%%%%%%
%%%%%%%%%%%%%%%%%%%%%%%%%%%%%%%%
\begin{figure*}
\centering
\includegraphics[width=0.30\textwidth]{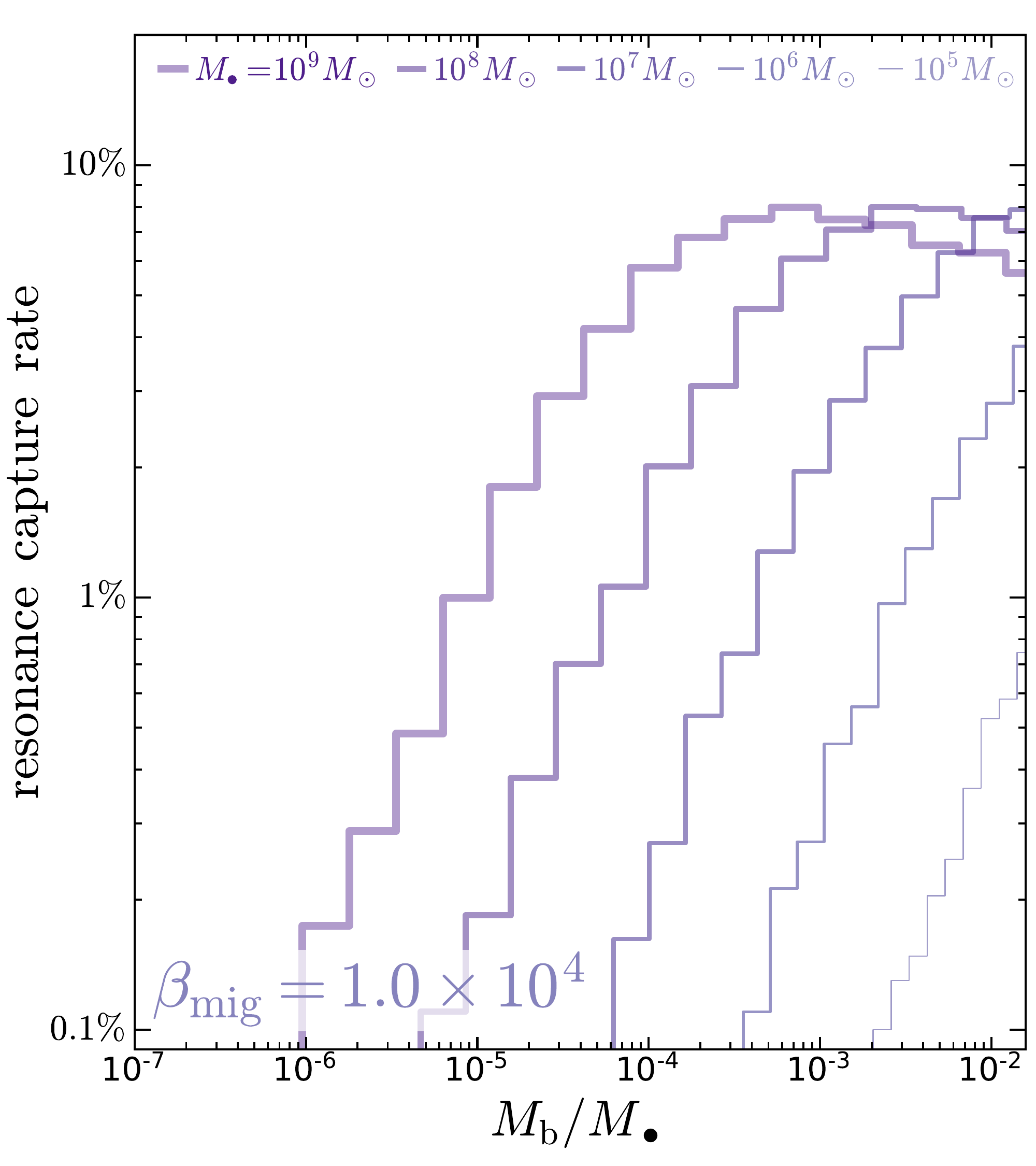}
\includegraphics[width=0.30\textwidth]{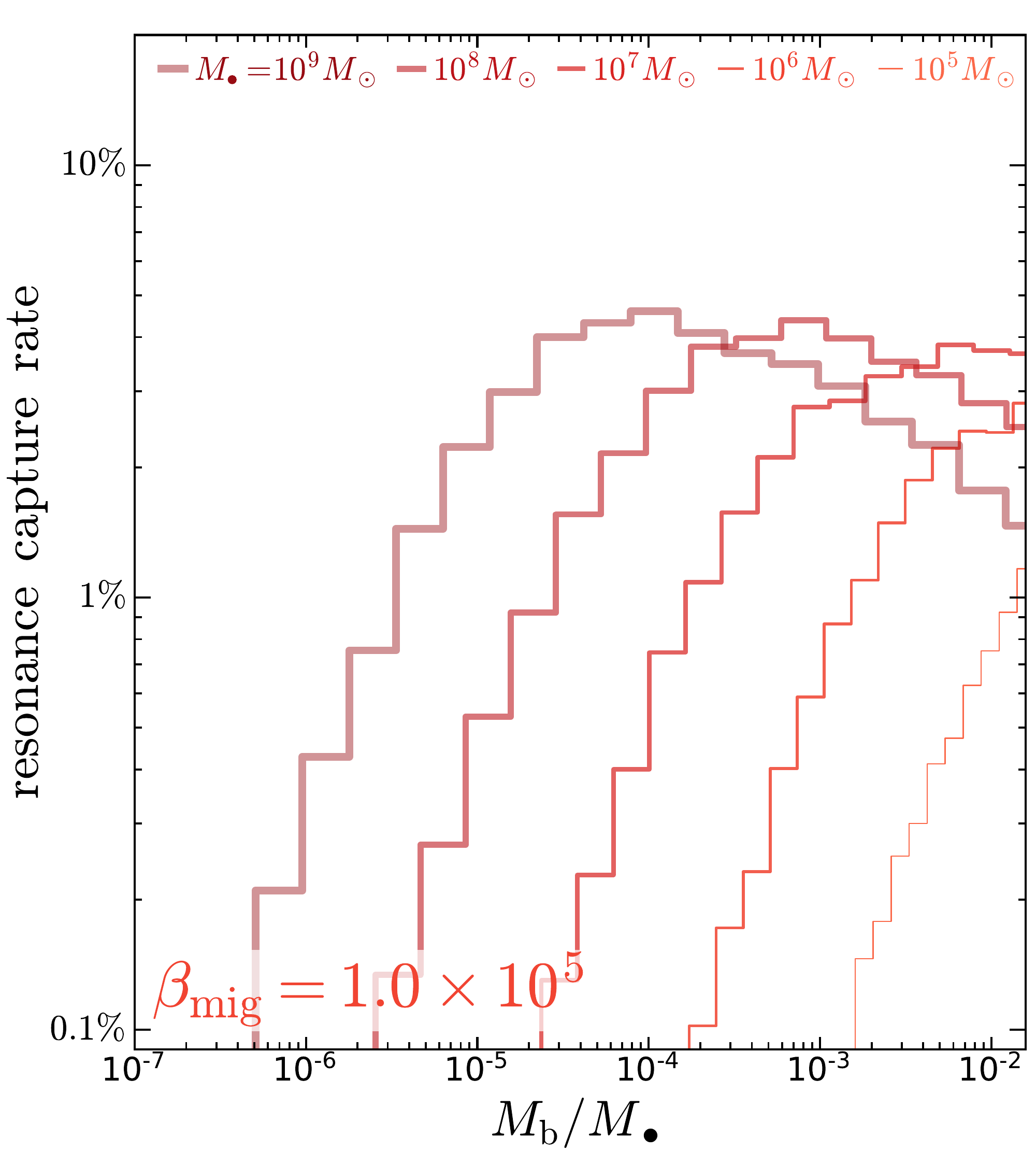}
\includegraphics[width=0.30\textwidth]{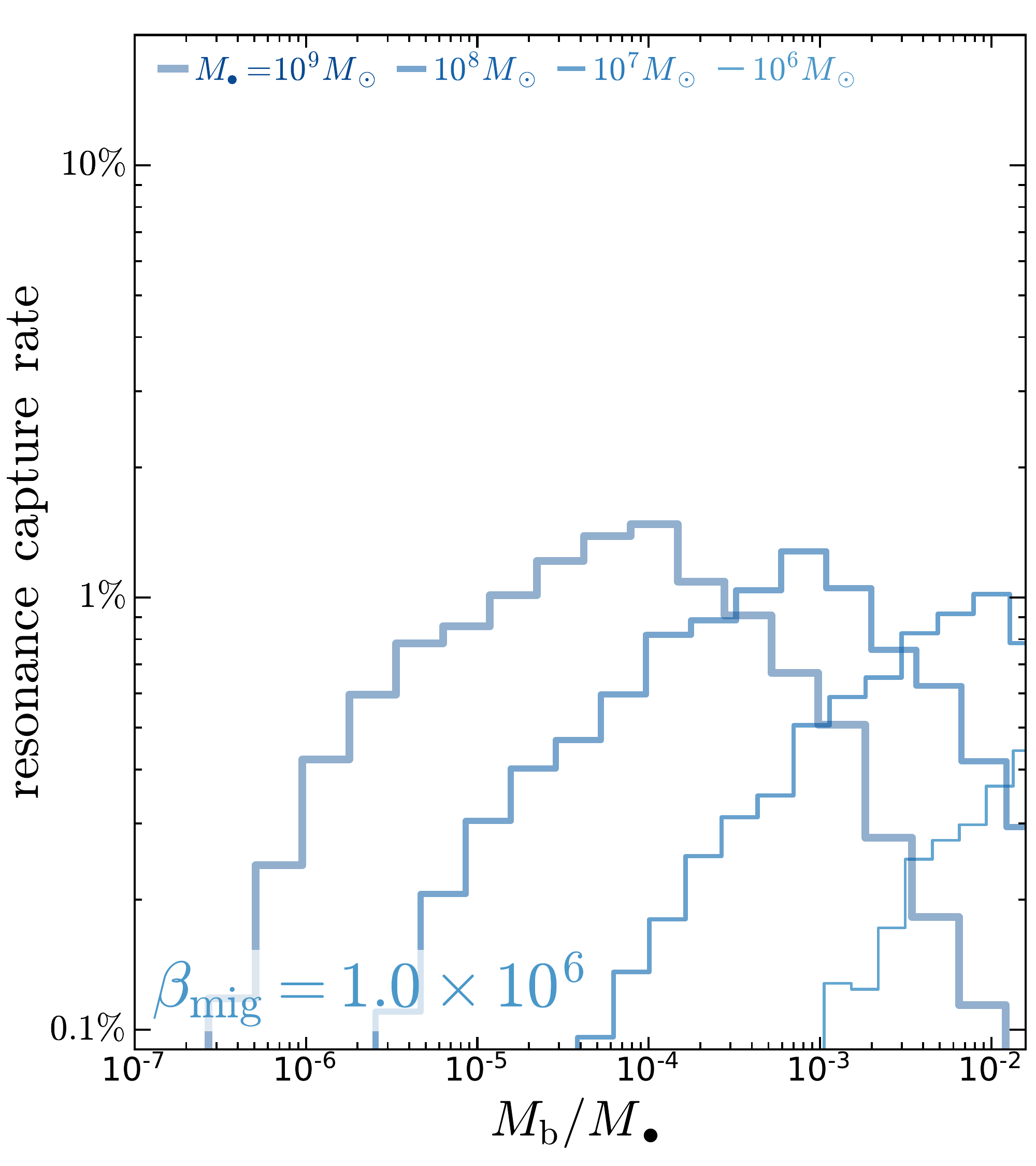}
\caption{Resonance capture rates. Similar to Figure~\ref{fig:crossing_rate}, but now requiring condition (i) (Section~\ref{sec:conditions}) to be satisfied, which states that
$\delta'_\star$ (Equation~\ref{eq:delta_prime_star}) is positive yet smaller than 0.25. In general, this additional requirement severely depress the captures rates relative to
 the crossing rates. At high BHB-to-SMBH mass ratios ($\gtrsim10^{-3}$), however, requiring capture is nearly equivalent to requiring crossing, and such configurations are more prolific in producing evection-accelerated mergers. Conversely, low BHB-to-SMBH mass ratios  ($\lesssim10^{-6}$) exhibit vanishingly small capture rates. As a consequence
 evection resonance capture of a binary composed of two stellar-mass black holes is virtually impossible.
\label{fig:capture_rate}}
\end{figure*}
%%%%%%%%%%%%%%%%%%%%%%%%%%%%%%%%
%%%%%%%%%%%%%%%%%%%%%%%%%%%%%%%%
%%%%%%%%%%%%%%%%%%%%%%%%%%%%%%%%%%%%%%%%%%%%%%%%%%%%%%%%%

In summary, intermediate mass-ratio BBHs containing an IMBH are the most likely objects to be captured into an evection resonance. As we show in the next section below, the final outcome of this resonance is an IMRI that is accelerated by the interceding evection dynamics.

\subsection{Long-term Behavior: Resonance Detuning and Evection-Accelerated Mergers}\label{sec:long_term}
In Figure~\ref{fig:resonance} we depict once again our fiducial example of Figure~\ref{fig:eccentricity} (right panel), this time over a timescale
of $6.7\times10^5$~yr, which the time it take for the BHB to merge.  The top panel depicts $e_{\rm b}$ and $e_{\rm b}^*$, showing that after resonant
capture these two curves follow each other closely up to a maximum eccentricity of $\simeq 0.5$. Once the eccentricity has grown significantly, the binary's hardening timescale shortens \citep[][Equation~\ref{eq:hardening2}]{pete64}, which in turn reverses the sign of 
of $\delta'$, now dominated by the rapid change in $a_{\rm b}$. The decrease in $\delta$ is evidenced by the turnover in the evolution of 
$e_{\rm b}^*\propto\sqrt{1+\delta}$ at $t\approx3.8\times10^5$~yr. As $\delta$ decreases further, the drift rate $|\delta'|$ increases, eventually breaking adiabatic invariance.
Adiabaticity breakdown ultimately leads to $\sigma$ going back to circulating
from $0$ to $2\pi$ at $t\approx4.7\times10^5$~yr (middle panel), which is sometimes referred to as the ``detuning'' of the resonance, explaining why 
$e_{\rm b}$ and $e_{\rm b}^*$ no longer track each other once the resonance has been detuned.

It may appear that, as the BHB is able to enter and then leave the resonant regime, little evidence of resonant behavior is left for us to find.
But we may identify two imprints that are indicative of past or concurrent resonant evolution. First, the BHB can enter the GW detectability band while still eccentric (see Section~\ref{sec:eccentric_mergers} below), producing waveforms and characteristic strains different from its circular counterpart. Second,
 the hardening timescale can be significantly reduced in relation to that of a quasi-circular orbit. Such  ``evection-acceleration'' of the merger is illustrated in the bottom panel of Figure~\ref{fig:resonance}, which compares the evolution of $a_{\rm b}$ through resonance capture and detuning (thick line) to that of an equivalent BHB evolving in isolation,  represented by  $a_{\rm b,circ}$ (thin line, Equation~\ref{eq:ab_circ}). The merger is evection-accelerated by a factor of about $2.5$, which can be significant in systems where BHBs would otherwise migrate across the entire AGN disk before they can merge.

%%%%%%%%%%%%%%%%%%%%%%%%%%%%%%%%%%%%%%%%%%%%%%%%%%%
%%%%%%%%%%%%%%%%%%%%%%%%%%%%%%%%%%%%%%%%%%%%%%%%%%%
\section{Evection Resonances in AGN Disks}\label{sec:agn_disks}
\subsection{Black Hole Binary Parameters in AGN}\label{sec:parameters}
We use Section~\ref{sec:conditions} to identify those binaries drawn from a random population of BHBs that should cross the evection commensurability. We generate a population of BHBs
by fixing the values of $M_\bullet$, $M_{\rm b}$ and $\beta_{\rm mig}$,   then
drawing $a_{\rm b}$ from a log-uniform distribution in $[100r_g,0.5 R_{\rm H}]$ (where $R_{\rm H}=(\tfrac{1}{3}M_{\rm b}/M_\bullet)^{1/3}$ is the BHB's Hill radius), and
drawing $q_{\rm b}$ from a log-uniform distribution in $[10^{-4},1]$ with the additional requirement that
$m_2=q_{\rm b}M_{\rm b}/(1+q_{\rm b})\geq 8M_\odot$.
The distance $ a_{\rm out}$ is  generated assuming that the BHB number density tracks the disk  surface density $\Sigma_{\rm disk}\propto r^{-\alpha}$,
and thus, we draw $a_{\rm out}$ values from a PDF $\propto a_{\rm out}^{1-\alpha}$ (we adopt $\alpha=3/2$; \citealp{sirk03}). 

We show the crossing
rates in Figure~\ref{fig:crossing_rate} for a variety of combinations of $M_\bullet$, $M_{\rm b}$ and $\beta_{\rm mig}$. The rates vary greatly, but, in general, faster migration timescales (smaller $\beta_{\rm mig}$) tend to produce more crossings. From those systems guaranteed to cross the commensurability, we can identify those expected to be captured into resonance by satisfying criteria (i), (ii) and (iii) of Section~\ref{sec:conditions}. These capture rates are depicted in Figure~\ref{fig:crossing_rate}. While, in general, there is a drop in rates when going from Figure~\ref{fig:crossing_rate} to Figure~\ref{fig:capture_rate}, some parameters produce near-equal rates of captures and crossing, with these being of order a few percent. Note that in all cases, the capture rates for stellar-mass BHBs is vanishingly small, and thus producing evection-accelerated eccentric mergers for $M_{\rm b}\sim{\cal O}(10M_\odot)$ is virtually impossible, at least for the simplified AGN disks and migration models that we assume here.

%%%%%%%%%%%%%%%%%%%%%%%%%%%%%%%%
%%%%%%%%%%%%%%%%%%%%%%%%%%%%%%%%
\begin{figure*}
\centering
\includegraphics[width=0.49\textwidth]{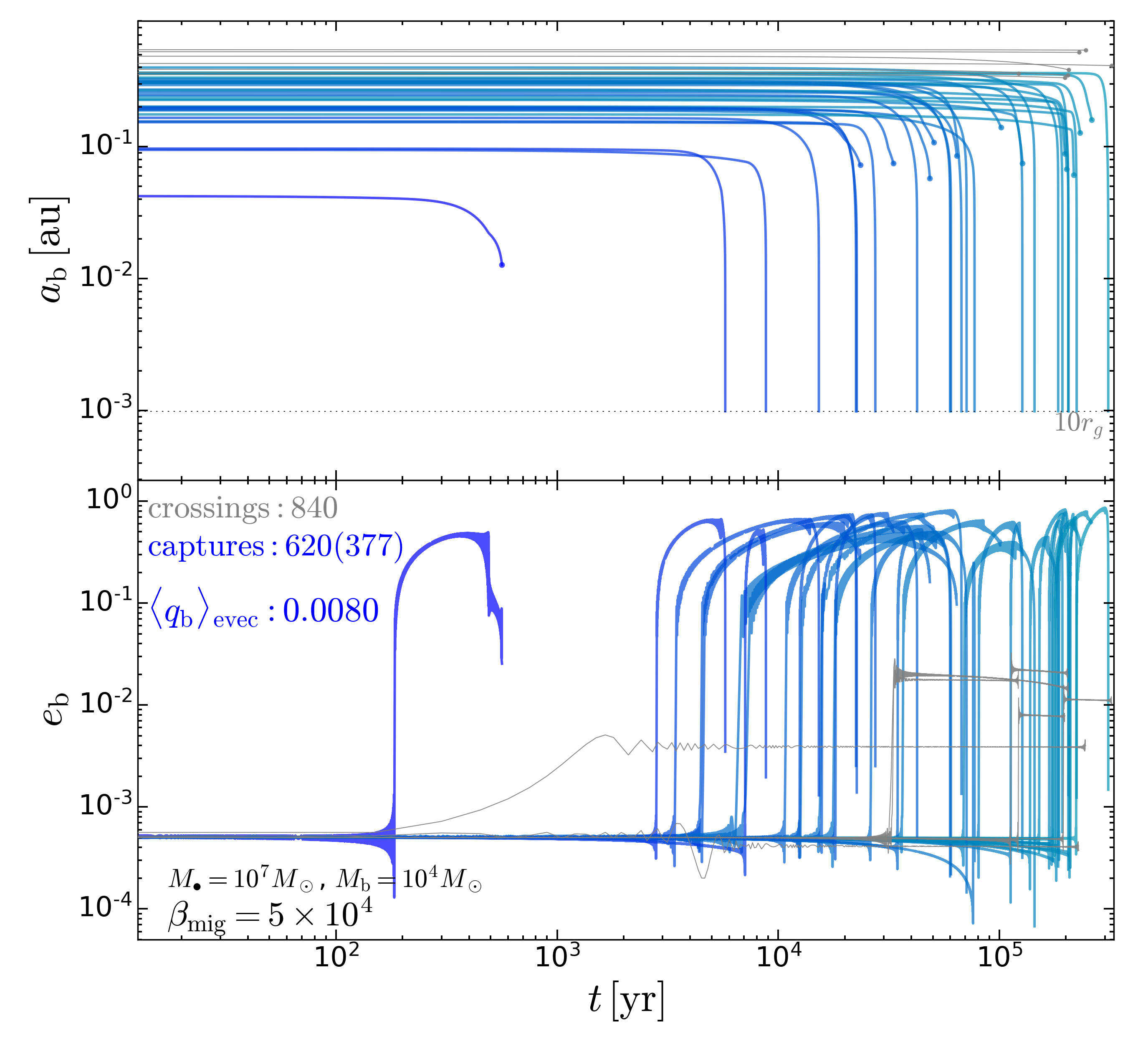}
\includegraphics[width=0.49\textwidth]{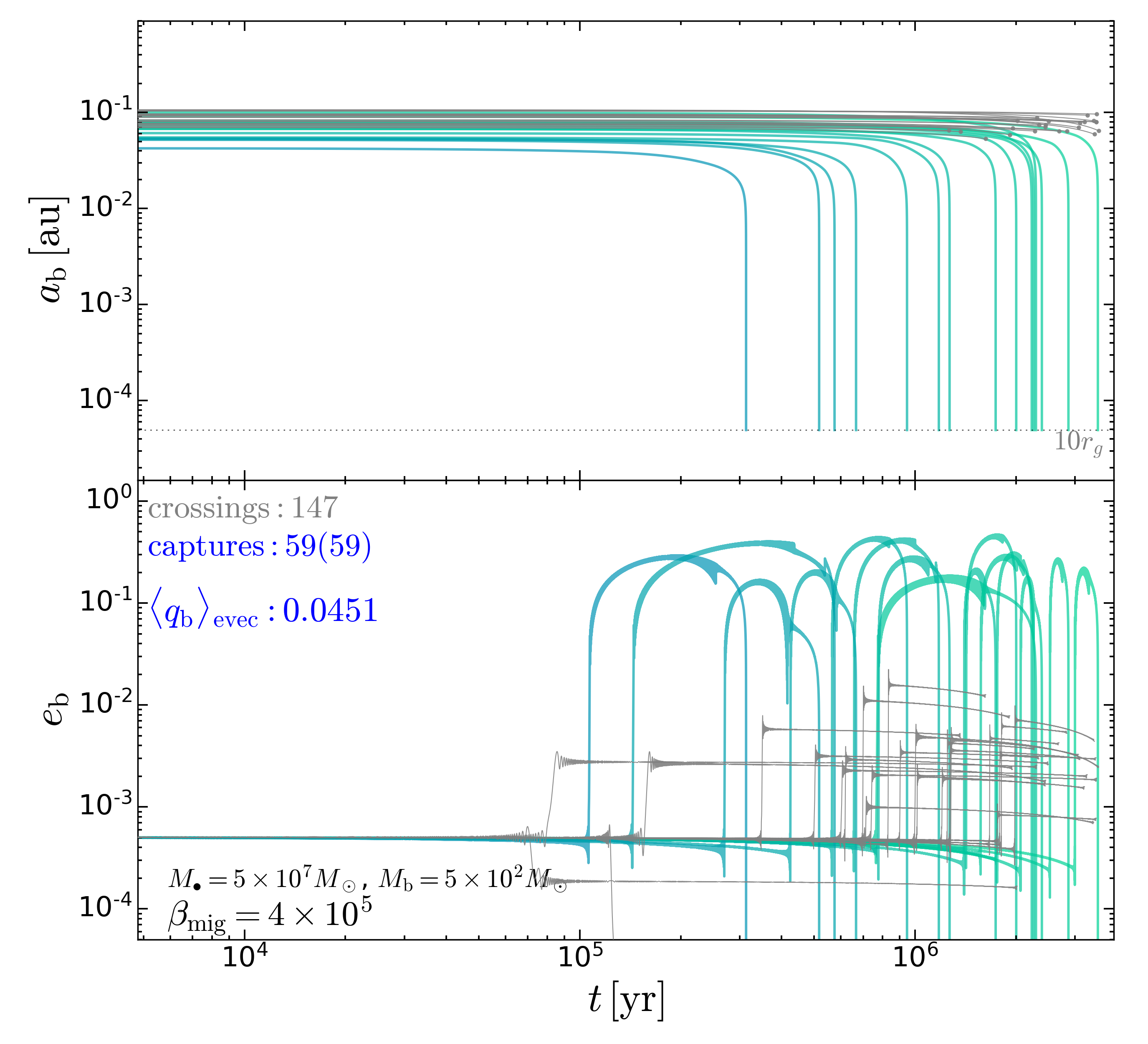}
\caption{Numerical solution of the equations of motion (\ref{eq:milankovitch_eom}) for a collection of  systems 
that cross the commensurability~(\ref{eq:commensurability}). The left column shows the evolution of $a_{\rm b}$ (top) and $e_{\rm b}$ (bottom) for 840 systems
with randomly generated values of $q_{\rm b}$, $a_{\rm b,0}$ and $a_{\rm out,0}$, with  fixed parameters
 $M_\bullet=10^7M_{\odot}$, $M_{\rm b}=10^4M_\odot$ and $\beta_{\rm mig}=5\times10^4$ ($5\%$ of system shown for clarity).
 The gray lines depict the BHBs that merely cross the evection commensurability (a $4.2\%$ rate),  and the blue lines depict those BHBs that were captured into resonance (a $3.1\%$ rate).  The separation $a_{\rm b}$ of non-resonant BHBs barely changes before being
 tidally disrupted by the SMBH (disruption is depicted by a filled circle). On the other hand, the separation of resonant BHBs decreases in all cases, and in 377 instances it decreases toward merger (top panel). The non-resonant and resonant population are distinguished by the evolution of $e_{\rm b}$ (bottom), which undergoes a one-time-only jump in the former case, while in the latter, eccentricity can grow until resonance detuning takes place (see Figure~\ref{fig:resonance}). For clarity, the ``trivial'' population of quasi-circular BHBs that will merge before migrating through the disk (purple region in Figure~\ref{fig:parameters}) is not shown. The mean secondary mass for resonant systems
 is $\langle M_2\rangle_{\rm evec}\approx M_{\rm b}\langle q_{\rm b}\rangle_{\rm evec}\approx 80M_\odot$.
The right column show the same as the left column but for parameters  $M_\bullet=5\times10^7M_{\odot}$, $M_{\rm b}=500M_\odot$ and $\beta_{\rm mig}=4\times10^5$ ($30\%$ of
systems shown for clarity).
In this case, only 145 systems cross the commensurability  (a $0.7\%$ rate) , out of which 57  are captured into resonance (a $0.3\%$ rate). The entire resonant population is able
to coalesce toward merger. The mean secondary mass for resonant systems is
 is $\langle M_2\rangle_{\rm evec}\approx  20M_\odot$.
\label{fig:population}}
\end{figure*}
%%%%%%%%%%%%%%%%%%%%%%%%%%%%%%%%
%%%%%%%%%%%%%%%%%%%%%%%%%%%%%%%%
%%%%%%%%%%%%%%%%%%%%%%%%%%%%%%%%%%%%%%%%%%%%%%%%%%%%%%%%%

%%%%%%%%%%%%%%%%%%%%%%%%%
\subsection{Numerical Integrations}
As a concrete example, we generate an $N{=}20000$ sample of coplanar, quasi-circular ($e_{\rm b,0}=0.0005$) BHBs 
 as described above. We adopt the parameters $M_\bullet=10^7M_\odot$, $M_{\rm b}=10^4M_\odot$ and $\beta=5\times10^4$.
According to the rate estimates of Section~\ref{sec:conditions}, 
this random sample contains 840 BHBs ($4.21\%$)  that
 will cross the commensurability. Likewise,
 the capture criterion indicates that $\simeq616$ BHBs ( $3.08\%$)
 should be captured into resonance. We evolve these 840 BHBs by numerically solving Equations~(\ref{eq:eom_milankovitch}) until either
 $a_{\rm b}=10 r_g$,  $a_{\rm b}=R_{\rm H}$  or $a_{\rm out}=0.01 a_{\rm out,0}$.
  The remaining sample of 19331 BHBs ($96.65\%$) evolve simply according to 
 Equation~(\ref{eq:ab_circ}), and consequently, we do not integrate those numerically. 
 The evolution of $a_{\rm b}$ and $e_{\rm b}$ for the  840 synthetic BHBs  is shown the left column of  Figure~\ref{fig:population} (only $5\%$ of curves are plotted).
In excellent agreement with the expectations, out of 840 BHBs,  220 merely cross the commensurability
without true capture (gray lines), while 620 BHBs are captured into evection resonance (blue lines). All these evection-captured BHBs experience a significant decrease
in separation, in contrast to their non-resonant  counterparts, which barely harden before being disrupted (when $a_{\rm b}\gtrsim R_{\rm H}$) as a consequence of  migration. Note, however, that not every evection-captured
BHBs is allowed to fully complete its merger. Indeed, 377 BHBs will reach zero separation in a finite time, while 243 are tidally disrupted by the SMBH while still eccentric (filled circles denote disruption). More interesting is the mass ratio $q_{\rm b}$ of those BHBs that are captured into resonance. The subset of 620 binaries captured into
evection have a mean mass ratio of $\langle q_{\rm b}\rangle_{\rm evec}=0.008$, i.e., the average secondary mass is $\langle M_2\rangle_{\rm evec}\simeq 80M_{\odot}$, or in other words, a stellar-mass black hole.

We repeat this experiment for lower black hole masses, 
adopting the parameters
$M_\bullet=5\times10^7M_\odot$, $M_{\rm b}=500M_\odot$ and $\beta=4\times10^5$, as shown in the right panel of Figure~\ref{fig:population}. This configuration is less fruitful for evection dynamics, resulting in only 147 ($0.735\%$) evection crossing, of which 58  ($0.29\%$)  are expected to be captured into resonance. Numerical integration confirms these estimates, producing
59 evection captures, all of which proceed toward merger without being tidally disrupted. In this example, we find $\langle q_{\rm b}\rangle_{\rm evec}=0.045$, or $\langle M_2\rangle_{\rm evec}\simeq 20M_{\odot}$, once again, a stellar-mass black hole.

%%%%%%%%%%%%%%%%%%%%%%%%%%%%%%%%
%%%%%%%%%%%%%%%%%%%%%%%%%%%%%%%%
\begin{figure*}
\centering
\includegraphics[width=0.49\textwidth]{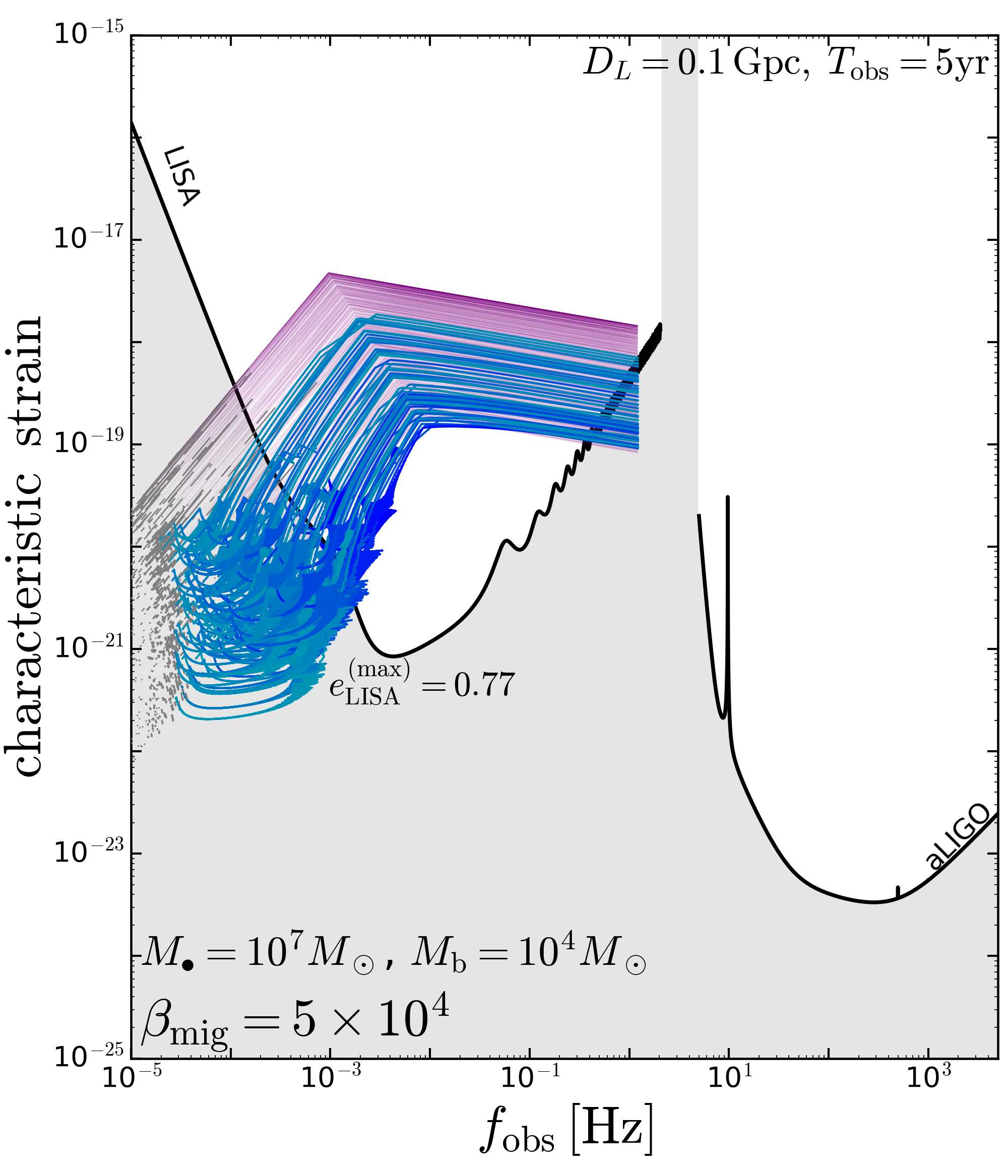}
\includegraphics[width=0.49\textwidth]{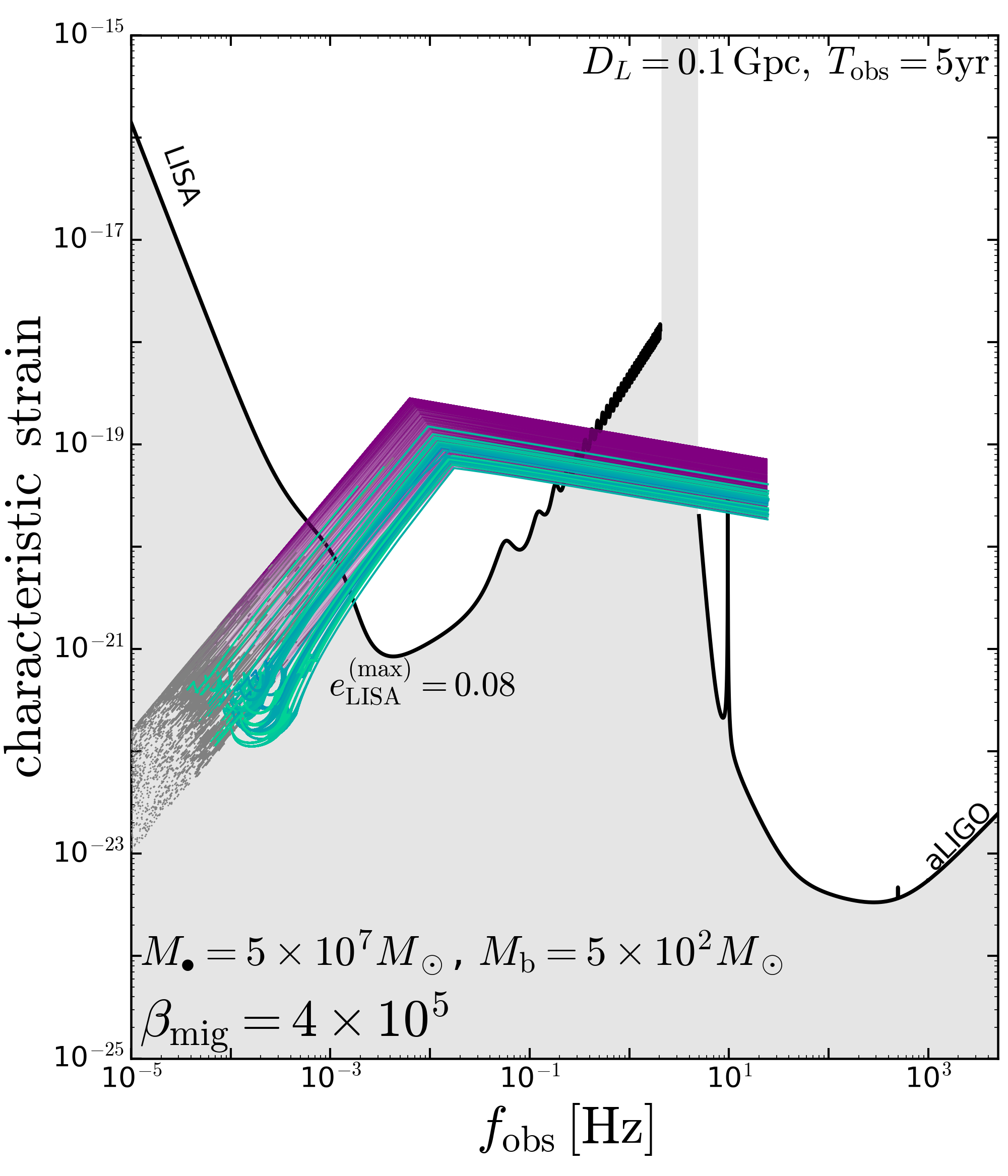}
\caption{Characteristic strain tracks for the synthetic population of BHBs show in Figure~\ref{fig:population}. In the left panel,
we show $[h_{c,{\rm peak}}^{2}]^{1/2}$ (Equation~\ref{eq:characteristic_strain})
for the 20000 BHBs generated with  $M_\bullet=10^7M_{\odot}$, $M_{\rm b}=10^4M_\odot$ and $\beta_{\rm mig}=5\times10^4$ (only 20\% of curves shown for clarity).
Evection-captured systems are shown in blue as in Figure~\ref{fig:population}. Non-resonant, evection-crossing systems. For completeness, we also include the 
population of binaries that evolve quasi-circularly according to Equation~(\ref{eq:ab_circ}). These include wide binaries that fail to coalesce (also in gray) and
and compact binaries that merge before being tidally disrupted (in purple). We plot the characteristic strain down to separations of $a_{\rm b}=3r_g$, beyond which the quadrupole formula is inaccurate. The resonant BHBs (promoted into to detectability by evection)
are able to enter the LISA band and, in many case, merge, solely due to their resonantly-excited eccentricities. Their tracks depart from the broken-power laws that are characteristic of quasi-circular coalescence 
\citep{sesa05}. The eccentricity of these systems at the moment of entering the LISA band is on average 0.43, with a maximum of 0.77.
In the right panel, we show the character strain tracks for  20000
BHBs generated with  $M_\bullet=5\times10^7M_{\odot}$, $M_{\rm b}=500M_\odot$ and $\beta_{\rm mig}=4\times10^5$ (all curves shown). In this case, BHBs enter the LISA
band with lower eccentricities (mean of 0.04 and maximum of 0.1) but, being of lower mass, are able to enter the LIGO band during their final plunge.
\label{fig:strains}}
\end{figure*}
%%%%%%%%%%%%%%%%%%%%%%%%%%%%%%%%
%%%%%%%%%%%%%%%%%%%%%%%%%%%%%%%%
%%%%%%%%%%%%%%%%%%%%%%%%%%%%%%%%%%%%%%%%%%%%%%%%%%%%%%%%%

%%%%%%%%%%%%%%%%%%%%%%%%%
\subsection{Gravitational Waves from Eccentric, Accelerated Black Hole Mergers}\label{sec:eccentric_mergers}

In Section~\ref{sec:long_term} we argued that the long-term net effect of the evection resonance, in addition to accelerating BHB coalescence, was  to change the GW signature of such mergers. Now, we study quantitatively how evection produces a detectable GW imprint. We do this
by computing, at each point in the binary's path toward coalescence, the GW characteristic strain (\citealp{MTW1973}, \S35.15; \citealp{thor89}) at peak emission $\tilde{h}_{c,{\rm peak}}$, and the dominant emitting frequency 
$f_{\rm peak}$. We plot $\tilde{h}_{c,{\rm peak}}$  as a function 
observed frequency $f_{\rm obs}=f_{\rm peak}(1+z)^{-1}$  in Figure~\ref{fig:strains} 
for the same set of integrations shown in Figure~\ref{fig:population}, and assuming that the AGN is located at luminosity distance of $D_L=0.1$~Gpc ($z=0.023$) and that the observing time is $T_{\rm obs}=5$~yr. Every `characteristic strain track' starts off with the spectral
dependence $\tilde{h}_{c,{\rm peak}}\propto f_{\rm obs}^{7/6}$ 
\citep[e.g.][]{sesa05}, but tracks quickly departure from this shape if they become resonant. Resonant BHB orbits are quickly shrunk, which allows them to enter the LISA band before being tidally disrupted by the SMBH. Moreover, at the time of entering the LISA band ($f_{\rm obs}\sim10^{-4}$Hz) the resonant BHBs still possess moderate-to-high eccentricities. For
the example in the left panel, the mean eccentricity when entering the LISA band is $\bar{e}_{\rm LISA}\approx0.43$, and the maximum is ${e}_{\rm LISA}^{\rm (max)}\approx0.77$. The high mass of the BHB ($M_{\rm b}=10^4 M_\odot$) means that the final merger is undetectable in the LIGO band, except perhaps for the ring-down stage. A smaller BHB, on the other hand, might be detectable by both LISA and LIGO. We illustrate this behavior in the right panel of Figure~\ref{fig:strains}. For $M_{\rm b}=500 M_\odot$, the merger
sweeps both LISA and LIGO bands. The LISA eccentricities, however, are significantly lower, with  $\bar{e}_{\rm LISA}\approx0.04$ ${e}_{\rm LISA}^{\rm (max)}\approx0.1$. And at even lower masses ($M_{\rm b}=500 M_\odot$), time spent in the LIGO band is increasingly longer, and $\bar{e}_{\rm LISA}$ get progressively smaller. Note however, that the price to pay to observe coalescence in both bands is steep: as shown in Section~\ref{sec:parameters}, BHBs with $M_{\rm b}\sim{\cal O}(100 M_\odot)$ have a have very small chance of being {\it captured} into evection resonance. And while evection crossing can indeed be far more likely, the eccentricity growth associated to crossing is rarely significant enough to be measurable.

%%%%%%%%%%%%%%%%%%%%%%%%%%%%%%%%%%%%%%%%%%%%%%%%%%%
%%%%%%%%%%%%%%%%%%%%%%%%%%%%%%%%%%%%%%%%%%%%%%%%%%%
\section{Discussion}\label{sec:discussion}
In this work, we have applied the evection resonance capture mechanism \citep[e.g.,][]{tou98} as a  purely dynamical channel  to produce eccentric BHB mergers in AGN disks.
 Key features of the evection mechanism are that it is of zeroth-order in inclination, and does not require high initial inclinations to operate, and that in principle, it can pump eccentricity to arbitrarily high values until GW damping overcomes resonant excitation. As such, this
 mechanism differs from other eccentricity-driving channels presented in the literature, such as GW capture \citep[e.g.,][]{sams20,taga21}, the secular ZLK mechanism  and its multiple variants \citep{petr17b,liu18,hoang18},
 or other linear secular resonances in the low-inclination limit \citep[e.g.,][]{ford00a,liu15b}.

%%%%%%%%%%%%%%%%%%%%%%%%%%%%%%%%%%%%%%%%%%%%%%%%%%%
\subsection{Formation Channels and Likelihood of Resonant Eccentricity Growth}\label{sec:formation_channels}
We have shown that the most likely binary to be captured into an evection resonance is one that is composed of an IMBH and a stellar-mass black hole.
Therefore, we need a sufficiently numerous population of both stellar-mass and intermediate-mass compact objects to coexist inside AGN disks.  

On the one hand, it is thought that stellar-mass black holes and neutron stars may form {\it in situ\/} through fragmentation of Toomre-unstable AGN zones \citep{thom05, ston17, gilb22}, or alternatively, may be captured from a pre-existing nuclear star cluster that predates any particular AGN episode \citep{syer91, bart17, pan21}.  Both of these mechanisms suffer from large theoretical uncertainties.  {\it In situ} formation is a complicated radiation hydrodynamics problem, and even after massive progenitor stars have formed the disk, their evolution may differ strongly from that of field stars \citep{ditt21}. And although  capture of a pre-existing compact object population through gas drag is less physically uncertain,  the end result of this process depends on the original inventory of compact objects in galactic nuclei, which is quite unclear.

On the other hand, there is the existence of IMBHs, which is even more uncertain. There are, however, good reasons to think that these may exist, perhaps in substantial numbers, in AGN disks.  First, if sustained super-Eddington accretion is possible, then embedded stellar-mass compact objects will rapidly grow to masses $\sim 10^{3-5} M_\odot$ in large portions of AGN parameter space \citep{good04}.  However, even in the absence of super-Eddington accretion, repeated mergers of stellar-mass BHs can lead to hierarchical growth and the production of IMBHs \citep{mcker12, mcker14}.  These repeated mergers rely on binary BH formation, which may happen due to gas-assisted BHB formation in generic regions of the disk \citep{taga20}, but may be especially favored in migration traps \citep{bell16}.

Given these broad uncertainties, we can remain cautiously optimistic about the possibility of AGN disks hosting binaries that consist of an IMBH and a stellar-mass black hole.

%%%%%%%%%%%%%%%%%%%%%%%%%%%%%%%%%%%%%%%%%%%%%%%%%%%
\subsection{Caveats and Extensions}\label{sec:caveats}
An important caveat to our work concerns the migration timescale
coefficient $\beta_{\rm mig}$ (Equation~\ref{eq:beta_mig}).
While we have assumed that $\beta_{\rm mig}$ is a constant, in reality it encapsulates the physics of migration due to nebular tides
\citep{lin79,gol80,war86,war97}, which depends on both the properties of the ``migrator'' as well as of the background disk. For instance, for the so-called Type-I migration, we have \citep[e.g.,][]{tana02}
\begin{equation}
\beta_{\rm mig}\sim \frac{M_\bullet^2}{M_{\rm b}\Sigma_{\rm disk}r_{\rm out}^2}h^2
\end{equation}
where $h^2$ and  $\Sigma_{\rm disk}$ are the disk aspect ratio and surface density, respectively. If we assume $M_\bullet=10^7M_\odot$
and $M_{\rm b}=10^4M_\odot$ (Figure~\ref{fig:parameters2}) and use 
the AGN disk models of \citet{sirk03} (their figure 2)
evaluated at  $r_{\rm out}\simeq2\times 10^3({\cal G}M_{\bullet}/c^2)\approx200$~au 
($h\simeq0.01$, $\Sigma_{\rm disk}\simeq10^6 {\rm gr}~{\rm cm}^{-3}\approx 10^{-1}M_\odot~{\rm au}^{-3}$),
we have that $\beta_{\rm mig}\approx5\times10^4$, which is in good
agreement with the value assumed for the model in Figure~\ref{fig:parameters2}.  However, if one chooses instead
to evaluate the Sirko-Goodman model  at
$r_{\rm out}\simeq 2000$~au
($h\simeq0.03$, $\Sigma_{\rm disk}\simeq10^4 {\rm gr}~{\rm cm}^{-3}\approx 10^{-3}M_\odot~{\rm au}^{-3}$), then we 
obtain $\beta_{\rm mig}\approx4.5\times10^6$.  Such
realistic disks can also exhibit non-monotonic profiles in $h^2$ and  $\Sigma_{\rm disk}$ \citep[see also][]{thom05}, which may result in 
``migration traps'' \citep{bell16}, which could concentrate in specific locations of the disk where IMBH binaries form, as well as where the evection-accelerated merger take place.
In the end, it is difficult to assess with certainty whether realistic AGN disks ultimately increase or decrease the likelihood of merger acceleration due to evection capture. A comprehensive answer may require a full-fledged population synthesis calculation, which is beyond the scope of this work.

An additional caveat is that we have ignored non-trivial hydrodynamical effects at the Hill sphere-scale  \citep[e.g.,][]{baru11}, which are expected to affect the orbital elements of the BHB.  The details, however, of this orbital evolution, remain highly uncertain and controversial, and are subject of active computational research \citep[e.g.,][]{li22,demp22}. And while binary contraction appears
to be the favored outcome of this configuration, there is some recent evidence pointing out to binary expansion as well.  According to
\citep{demp22}, binary expansion can happen when $a_{\rm b}\ll R_{\rm H}$, i.e., when the BHB may be able to form--and accrete from--a circumbinary disk within its own Hill Sphere. If this is the case, such a result would be consistent with earlier work reporting that steady-state accretion onto binaries from coplanar circumbinary disks lead to binary expansion \citep{mir17,mun19,moo19}.

In addition, the mere fact that the binary is increasing its mass can significantly alter the picture laid out in Section~\ref{sec:conditions}, since now $\dot{M}_{\rm b}\neq0$ enters
into Equation~(\ref{eq:delta}) via $n_{\rm b}$ and $r_g$.
When binary accretion is taken into account, the drift rate (Equation~\ref{eq:delta_prime_2})
is
modified to:
\begin{equation}
\begin{split}
\delta'=\frac{4}{75}\frac{n_{\rm b}^2}{n_{\rm out}^3}
&\bigg[-\frac{\dot{a}_{\rm b}}{a_{\rm b}}\left(1-8\frac{n_{\rm b}}{n_{\rm out}}\frac{r_g}{a_{\rm b}}\right)\\
&
+\frac{\dot{M}_{\rm b}}{M_{\rm b}}\left(\frac{1}{3}-4\frac{n_{\rm b}}{n_{\rm out}}\frac{r_g}{a_{\rm b}}\right)
\\
&
+\frac{\dot{a}_{\rm out}}{a_{\rm out}}\left(1-6\frac{n_{\rm b}}{n_{\rm out}}\frac{r_g}{a_{\rm b}}\right)
\bigg]
\end{split}
\end{equation}
from which it can be seen that mass accretion can contribute 
a positive amount to the drift rate, potentially expanding the parameters for which $\delta'>0$, or perhaps speeding up the drift rate as to make resonance capture impossible. Moreover, if circumbinary accretion contributes non-trivially to the evolution $\dot{a}$, the drift rate can be once again significantly altered. These extensions are beyond the scope of this present paper but can be explored in future work.

%%%%%%%%%%%%%%%%%%%%%%%%%%%%%%%%%%%%%%%%%%%%%%%%%%%
%%%%%%%%%%%%%%%%%%%%%%%%%%%%%%%%%%%%%%%%%%%%%%%%%%%
\section{Summary}\label{sec:summary}
In this work, we have presented a new application of nonlinear resonance capture to the problem of a BHB embedded in an AGN disk. Our findings are as follows:
\begin{itemize}
    \item[(i)] The tidal field of the central SMBH can   influence the evolution of the BHB when a 1:1 commensurability is reached between the binary's apsidal precession rate due to GR, $\dot{\omega}_{\rm GR}$, and the orbital frequency $n_{\rm out}$. When the commensurability is maintained in time, the system is said to be captured in an {\it evection resonance}, which results in arbitrary growth of the BHB eccentricity. 
    \item[(ii)] For evection resonance to be possible, the BHB must migrate in the AGN disk. Migration must be faster than the binary's hardening, yet slow enough to permit the capture. The stringent conditions for resonance capture make the evection resonance of a binary with two stellar-mass black holes virtually impossible.
       \item[(iii)] A resonant BHB experiences the significant acceleration of its 
    GW-driven coalescence, owing to the shrinking of the pericenter separation. This acceleration increases the merger rates in the AGN disk, since most of the BHBs that become resonant would be tidally disrupted by the SMBH as a consequence of disk migration.
    \item[(iv)] Evection-acceleration is most efficient for BHBs consisting of a stellar mass black hole and an IMBH, meaning that the vast majority of the evection-accelerated mergers would emit GWs in the LISA band as an IMRI.

     \item[(v)] BHBs with masses in the range $10^3-10^4~M_\odot$ could be observable entering the LISA band with significant eccentricities ($\gtrsim0.6$). Smaller BHBs ($100-1000~M_\odot$) could enter the LISA band at smaller, yet  still non-negligible eccentricities ($\lesssim0.1$) and also enter the LIGO band before  final plunge. The latter systems, however, are much less likely to be captured into evection resonance than the former.

\end{itemize}

\paragraph{Notes}
We acknowledge that, recently,
Bhaskar, Li \& Lin (2022) have also carried out a study on evection resonances of stellar mass black hole binaries in AGN disks.

%%%%%%%%%%%%%%%%%%%%%%%%%%%%%%%%%%%%%%%%%%%%%%%%%%%
%%%%%%%%%%%%%%%%%%%%%%%%%%%%%%%%%%%%%%%%%%%%%%%%%%%
%%%%%%%%%%%%%%%%%%%%%%%%%%%%%%%%%%%%%%%%%%%%%%%%%%%
%%%%%%%%%%%%%%%%%%%%%%%%%%%%%%%%%%%%%%%%%%%%%%%%%%%
%%%%%%%%%%%%%%%%%%%%%%%%%%%%%%%%%%%%%%%%%%%%%%%%%%%
%%%%%%%%%%%%%%%%%%%%%%%%%%%%%%%%%%%%%%%%%%%%%%%%%%%
%%%%%%%%%%%%%%%%%%%%%%%%%%%%%%%%
\acknowledgments{DJM thanks Yoram Lithwick
for helpful discussions on nonlinear resonances and Dong Lai and Yubo Su for discussing their work on evection dynamics. 
DJM acknowledges partial support from the Cottrell Fellowship Award from the Research Corporation for Science Advancement, which is partially funded by NSF grant CHE-2039044.
CP acknowledges support from ANIDMillennium Science Initiative-ICN12\_009, CATA-Basal AFB-170002, ANID BASAL project FB210003, FONDECYT Regular grant 1210425 and ANID + REC Convocatoria Nacional subvencion a la instalacion en la Academia convocatoria 2020 PAI77200076. FAR acknowledges support from NSF Grant AST-2108624 and NASA Grant 80NSSC21K1722.}

\bibliographystyle{apj}

%%%%%%%%%%%%%%%%%%%%%%%%%%%%%%%%%%%%%
%%%%%%%%%%%%%%%%%%%%%%%%%%%%%%%%%%%%%
\appendix
%%%%%%%%%%%%%%%%%%%%%%%%%%%%%%%%%%%%%
\section{Second Fundamental Model of Resonance}\label{app:fundamental}
To cast  ${\cal K}$ into the standard Henrard-Lemaitre form (or ``Andoyer form''; \citealp{ferrazmello2007}), we first assume low eccentricity ($\Sigma\ll \Lambda$) and  further simplify ${\cal K}$ by
retaining first order terms in $\Sigma/ \Lambda$ for the resonant part and
second order terms in $\Sigma/ \Lambda$ for the non-resonant part \citep[e.g.][]{tou98}:
\begin{equation}
\begin{split}
{\cal K} \approx &\bigg(n_{\rm out}-n_{\rm b}\frac{3r_g}{a_{\rm b}}
-\frac{n_{\rm out}^2}{n_{\rm b}}
\frac{3}{4}\bigg)\Sigma
+\bigg(\frac{n_{\rm out}^2}{n_{\rm b}}
\frac{3}{8}-n_{\rm b}\frac{3r_g}{a_{\rm b}}
\bigg)\frac{\Sigma^2}{\Lambda}-\frac{n_{\rm out}^2}{n_{\rm b}}\frac{15}{4}\Sigma\cos2\sigma
,\!\!\!\!\!\!
\end{split}
\end{equation}
where we have dropped irrelevant constants. Second, 
we
shift the canonical coordinate $\sigma\rightarrow \tfrac{\pi}{2}+\hat{\sigma}$, which is accompanied by the trivial transformation
$\Sigma\rightarrow \hat{\Sigma}=\Sigma$. 
\begin{equation}
\begin{split}
{\cal K} = &
\bigg(n_{\rm out}-n_{\rm b}\frac{3r_g}{a_{\rm b}}
-\frac{n_{\rm out}^2}{n_{\rm b}}
\frac{3}{4}\bigg)\hat{\Sigma}
-\bigg(
n_{\rm b}\frac{3r_g}{a_{\rm b}}
-\frac{3}{8}\frac{n_{\rm out}^2}{n_{\rm b}}
\bigg)\frac{\hat{\Sigma}^2}{\Lambda}
+\frac{n_{\rm out}^2}{n_{\rm b}}\frac{15}{4}\hat{\Sigma}\cos2\hat{\sigma}
,\!\!\!\!\!\!
\end{split}
\end{equation}
Next,  we take
the rescaling
\begin{equation}\label{eq:rescaling}
\tau = \frac{15}{4}\frac{n_{\rm out}^2}{n_{\rm b}}t,
\;\;\;\;\;
r=-\hat{\sigma},
\;\;\;\;\;
R=\left(\frac{2}{5}\frac{n_{\rm b}^2}{n_{\rm out}^2}
\frac{r_g}{a_{\rm b}}-\frac{1}{20}\right)
\frac{\hat{\Sigma}}{\Lambda}
\end{equation}
In order to preserve canonicity, coordinate and time rescalings of the type $t\rightarrow\tau=a t$,
$\hat{\sigma} \rightarrow r=b \hat{\sigma}$, and
$\hat{\Sigma}\rightarrow R=c \hat{\Sigma}$ must
be accompanied by a rescaling of the Hamiltonian in the form
to $\tilde{\cal K}=(bc/a){\cal K}$. Thus, we obtain the final Hamiltonian 
\begin{equation}\label{eq:resonant_hamiltonian}
\tilde{\cal K}
=-\left[-\frac{1}{5}+\frac{4n_{\rm b}}{15n_{\rm out}}\left(1-\frac{n_{\rm b}}{n_{\rm out}}\frac{3r_g}{a_{\rm b}}\right)\right]R
+2R^2-{R}\cos2r~.
\end{equation}
Identifying $\hat{\cal K}$ with the Henrard-Lemaitre Hamiltonian \citep[e.g.,][]{bor84}
$
H_k=(2k-5)(1+k\delta)R+kR^2-(3-k)^{1+k/2}R^{k/2}\cos kr
$
with $k=2$, we can define the drift parameter
\begin{equation}\label{eq:delta2}
\delta\equiv-\frac{3}{5}+\frac{2n_{\rm b}}{15n_{\rm out}}\left(1-3\frac{n_{\rm b}}{n_{\rm out}}\frac{r_g}{a_{\rm b}}\right)~.
\end{equation}
In Poincar\'e rectangular coordinates,
$\xi=\sqrt{2R}\sin r$ and $\eta=\sqrt{2R}\cos r$,
the Hamiltonian (\ref{eq:resonant_hamiltonian}) has stable equilibrium points
\begin{equation}
\xi_*=0~,\;\;\;\;\eta_*=\pm\sqrt{1+\delta},
\end{equation}
which, after reinstating the original $(\sigma,\Sigma)$
canonical pair, translates into the fixed points highlighted
in Figure~\ref{fig:hamiltonian}:
\begin{equation}
\sigma^\ast=\pm \frac{\pi}{2}~,
\;\;\;\;\;\;\;\;
\Sigma^*=
\Lambda\left[\frac{2}{5}\frac{n_{\rm b}^2}{n_{\rm out}^2}
\frac{r_g}{a_{\rm b}}-\frac{1}{20} \right]^{-1} \frac{1+\delta}{2}
\approx \Lambda\frac{5}{2}\frac{n_{\rm out}^2}{n_{\rm b}^2}
\frac{a_{\rm b}}{r_g} \frac{1+\delta}{2}~.
\end{equation}
%

%%%%%%%%%%%%%%%%%%%%%%%%%%%%%%%%%%%%%%%%%%%%%%%%%%%%%%%%%%%%%%%%%
\section{Characteristic Gravitational Wave Strain}\label{app:strains}
To get a sense of the ``dominant GW signal'' or ``characteristic strain'' associated to the BH binaries as they evolve in separation and eccentricity,
a common trick is to identify the Fourier harmonic emitting the most GW power \citep[][is a recent example of such approach]{krem19}. First, the waveforms $h_+(t)$ and $h_\times(t)$ of a GW event can be decomposed into  Fourier series as  $\sum \tilde{h}_{+ k}{\rm e}^{{\rm i }2\pi f_kt}$  and  $\sum \tilde{h}_{\times k}{\rm e}^{{\rm i}2\pi f_kt}$, respectively, where $f_k=kn_{\rm b}/(2\pi)$.  Decomposing the energy 
carried away by GW into a sum $\dot{E}_{\rm GW}=\sum_k \dot{E}_k$, it can be shown that
\begin{equation}\label{eq:gw_energy}
\dot{E}_k=2\pi^2D^2\frac{c^3}{\cal G} f_k^2 (|\tilde{h}_{+ k}|^2+ |\tilde{h}_{\times k}|^2)
\equiv \pi^2D^2\frac{c^3}{\cal G} f_k^2 \tilde{h}_k^2
\end{equation}
for $k\geq1$,(\citealp{MTW1973}, \S35.15; \citealp{thor89}), where $D$ is the distance to the source. For a Keplerian binary, \citet{pete63} derived a harmonic sum ($k\geq1$)
for $\dot{E}_{\rm GW}$ obtained using the quadrupole formula \citep[e.g.,][]{MTW1973}. Using elliptic expansions,  they find
\begin{equation}\label{eq:gw_binary}
\dot{E}_k=\frac{64}{5}n_{\rm b}\frac{\mu_{\rm b}}{M_{\rm b}}\left(\frac{a}{r_g}\right)^{-\frac{5}{2}}
\!  |E_{\rm b}|g(k,e_{\rm b})
\end{equation}
where $E_{\rm b}=-{\cal G}\mu_{\rm b}M_{\rm b}/(2a_{\rm b})$ is the binary's
total binding energy and $g(k,e_{\rm b})$ is the function defined in equation 20 of \citet{pete63}.  

Equating~(\ref{eq:gw_energy}) and ~(\ref{eq:gw_binary}), one can derive an expression for $\tilde{h}_k$. More importantly,
however, we are interested in deriving an expression for the {\it characteristic strain} $\tilde{h}_{c,k}\simeq \tilde{h}_k\sqrt{N_{\rm cyc}}$,
where $N_{\rm cyc}\equiv f_k^2/\dot{f}_k$ is the number of cycles the inspiral spends emitting at a frequency in the vicinity of $f_k$ \citep{thor95,flan98,finn00}.
Thus
\begin{equation}
\begin{split}
\tilde{h}_{c,k}^2&=\frac{2}{\pi^2 D^2}\frac{\cal G}{c^3}\frac{\dot{E}_k}{\dot{f}_k}\times \min[1,({\dot{f}_k}/{f_k})T_{\rm obs}]\\
\end{split}
\end{equation} 
\citep[e.g.,][]{bara04a}, where the correcting factor $\min[1,({\dot{f}_k}/{f_k})T_{\rm obs}]$ compensates for low-frequency signals that are effectively
stationary during an observation period $T_{\rm obs}$, and thus with a number of cycles is given by $N_{\rm cyc}\simeq f_k T_{\rm obs}$ \citep{cutl94,sesa05}.

The sweeping of frequencies is given by the orbital decay of the binary (Equation~\ref{eq:hardening2}),
i.e., $\dot{f}_k/f_k=-\tfrac{3}{2}\dot{a}_{\rm b}/a_{\rm b}$.
\begin{equation}\label{eq:characteristic_strain}
\begin{split}
\tilde{h}_{c,k}^2&=\frac{4}{3\pi k}\frac{\mu_{\rm b}}{M_{\rm b}}
\frac{a_{\rm b}^2}{D^2}\!
\left(\frac{a_{\rm b}}{r_g}\!\right)^{\!\!-\tfrac{3}{2}}
\!\!\frac{g(k,e_{\rm b})}{F(e_{\rm b})}
 \min[1,\!-\tfrac{3}{2}\frac{\dot{a}_{\rm b}}{a_{\rm b}}T_{\rm obs}]\\
\end{split}
\end{equation} 
Focusing on the peak of GW emission, we evaluate $\tilde{h}_{c,k}^2$ at  $k=k_{\rm peak}$, which \citet{wen03} empirically identified
as to satisfy:
\begin{equation}
k_{\rm peak}=\frac{2(1+e_{\rm b})^{1.1954}}{(1-e_{\rm b}^2)^{3/2}}
\end{equation} 
and consequently, corresponds to a peak emission frequency $f_{\rm peak}=k_{\rm peak} n_{\rm b}/(2\pi)$.

\end{document}